\documentclass[10pt,journal]{IEEEtran}

\ifCLASSINFOpdf

\else

\fi

\usepackage{mathrsfs}
\usepackage{amsmath}
\usepackage{amsfonts}
\usepackage{amsthm}
\usepackage{amssymb}
\usepackage{graphicx}
\usepackage{subfigure}
\usepackage{indentfirst}
\usepackage{array}
\usepackage{epstopdf}
\usepackage{cite}
\usepackage{enumerate}
\usepackage{bm}
\usepackage{dsfont}
\usepackage{multirow}
\usepackage{verbatim}
\usepackage{stfloats}
\usepackage{makecell}
\usepackage{cancel}
\usepackage{threeparttable}
\usepackage{tikz}
\usetikzlibrary{positioning}
\usetikzlibrary{spy}
\usepackage{wrapfig}
\usepackage{pgfplots}
\usepackage{nicefrac}
\usepackage{thmtools,thm-restate}
\usepackage{color}

\usepackage{booktabs}
\usepackage{graphicx}
\usepackage{bbm}
\usepackage{adjustbox}
\usepackage{multirow}
\usepackage{tabularx}
\usepackage[misc]{ifsym}
\usepackage{bbding}
\usepackage{colortbl}
\usepackage{algorithm}
\usepackage{algorithmicx}
\usepackage{algpseudocode}
\usepackage{xcolor}
\usepackage{hyperref}

\algnewcommand{\LineComment}[1]{\State {\color{gray}\(\triangleright\) #1}}
\definecolor{lightblue}{rgb}{0.9, 0.92, 0.96}
\definecolor{subsectioncolor}{rgb}{0, 0, 0.706}   
\definecolor{mylightblue}{rgb}{0.851, 0.898, 0.941} %
\definecolor{LightCyan}{rgb}{0.88,1,1}
\definecolor{LightOrange}{rgb}{1,0.85,0.70}

\usepackage{pifont}
\newcommand{\cmark}{\ding{51}}%
\newcommand{\xmark}{\ding{55}}%

\newcommand{\argmin}{\mathop{\mathrm{argmin}}\limits}

\definecolor{C0}{rgb}{0.121569, 0.466667, 0.705882}
\definecolor{C1}{rgb}{1.000000, 0.498039, 0.054902}
\definecolor{C2}{rgb}{0.172549, 0.627451, 0.172549}
\definecolor{C3}{rgb}{0.839216, 0.152941, 0.156863}
\definecolor{C4}{rgb}{0.580392, 0.403922, 0.741176}
\definecolor{C5}{rgb}{0.549020, 0.337255, 0.294118}
\definecolor{C6}{rgb}{0.890196, 0.466667, 0.760784}
\definecolor{C7}{rgb}{0.498039, 0.498039, 0.498039}
\definecolor{C8}{rgb}{0.737255, 0.741176, 0.133333}
\definecolor{C9}{rgb}{0.090196, 0.745098, 0.811765}
\definecolor{trolleygrey}{rgb}{0.5, 0.5, 0.5}

\begin{document}

\title{DiffCom: Channel Received Signal is a Natural Condition to Guide Diffusion Posterior Sampling}

\author{Sixian Wang,~\IEEEmembership{Member,~IEEE},
	Jincheng Dai,~\IEEEmembership{Member, IEEE},
	Kailin Tan,~\IEEEmembership{Graduate Student Member, IEEE},\\
	Xiaoqi Qin,~\IEEEmembership{Senior Member, IEEE},
	Kai Niu,~\IEEEmembership{Member, IEEE},
	and Ping Zhang,~\IEEEmembership{Fellow, IEEE}
	
	\thanks{This work was supported in part by the National Key Research and Development Program of China under Grant 2024YFF0509700, in part by the National Natural Science Foundation of China under Grant 62321001, Grant 62293481, Grant  62371063, Grant 92267301, and Grant 62201089. in part by the Beijing Municipal Natural Science Foundation under Grant L232047, and Grant 4222012, in part by Program for Youth Innovative Research Team of BUPT No. 2023YQTD02, in part by BUPT Excellent Ph.D. Students Foundation CX2023118. (\emph{Corresponding author: Jincheng Dai})}

	\thanks{Sixian Wang, Jincheng Dai, Kailin Tan, and Kai Niu are with the Key Laboratory of Universal Wireless Communications, Ministry of Education, Beijing University of Posts and Telecommunications, Beijing 100876, China. (email: sixian@bupt.edu.cn, daijincheng@bupt.edu.cn)}
	
	\thanks{Xiaoqi Qin and Ping Zhang are with the State Key Laboratory of Networking and Switching Technology, Beijing University of Posts and Telecommunications, Beijing 100876, China.}
	
	\vspace{0em}
}

\maketitle

\begin{abstract}
	End-to-end visual communication systems typically optimize a trade-off between channel bandwidth costs and signal-level distortion metrics. However, under challenging physical conditions, this traditional coding and transmission paradigm often results in unrealistic reconstructions with perceptible blurring and aliasing artifacts, despite the inclusion of perceptual or adversarial losses for optimizing. This issue primarily stems from the receiver's limited knowledge about the underlying data manifold and the use of deterministic decoding mechanisms. To address these limitations, this paper introduces \emph{DiffCom}, a novel end-to-end \emph{generative communication} paradigm that utilizes off-the-shelf generative priors and probabilistic diffusion models for decoding, thereby improving perceptual quality without heavily relying on bandwidth costs and received signal quality. Unlike traditional systems that rely on deterministic decoders optimized solely for distortion metrics, our \emph{DiffCom} leverages raw channel-received signal as a fine-grained condition to guide stochastic posterior sampling. Our approach ensures that reconstructions remain on the manifold of real data with a novel confirming constraint, enhancing the robustness and reliability of the generated outcomes. Furthermore, \emph{DiffCom} incorporates a blind posterior sampling technique to address scenarios with unknown forward transmission characteristics. Extensive experimental validations demonstrate that \emph{DiffCom} not only produces realistic reconstructions with details faithful to the original data but also achieves superior robustness against diverse wireless transmission degradations. Collectively, these advancements establish \emph{DiffCom} as a new benchmark in designing generative communication systems that offer enhanced robustness and generalization superiorities. Our project and open source code are available at: \url{https://semcomm.github.io/DiffCom}

\end{abstract}

\begin{IEEEkeywords}
	Generative communications, source and channel coding, diffusion models, perceptual quality.
\end{IEEEkeywords}

\IEEEpeerreviewmaketitle

\section{Introduction}\label{section_introduction}

\subsection{Background}

\IEEEPARstart{T}{raditional} communication systems are typically optimized under the source-channel separation framework, which separates the processes of representation, compression, and transmission of information. This approach utilizes rate-distortion theory for source coding and channel coding theory for transmission, aiming to minimize the size of the source data under a distortion constraint, such as mean-squared error (MSE), while ensuring reliable data transmission over noisy channels subject to the constraints of channel capacity \cite{cover1999elements}. This framework has been foundational in advancing multiple generations of digital communication systems.

Recent advancements in deep learning have encouraged the exploration of data-driven solutions for end-to-end communications, specifically through the concept of joint source-channel coding (JSCC). Early works in deep learning-based JSCC (DeepJSCC) typically implement an autoencoder architecture that directly maps source information to channel-input symbols, taking into account the impairments of wireless channels to maximize end-to-end application performance (e.g., minimizing distortion (such as MSE) of reconstructed images). These models have demonstrated performance comparable to, or sometimes superior to, traditional methods like BPG (compatible with HEVC intra coding) + LDPC for small-sized CIFAR-10 images \cite{djscc, xu2021wireless, yang2022ofdm, yang2024swinjscc}. 

Inspired by the success of nonlinear transform coding (NTC) in neural data compression \cite{balle2020nonlinear}, the nonlinear transform source-channel coding (NTSCC) approach integrates a learned entropy model of the intermediate encoder features to determine necessary communication rates for achieving specified distortion levels, adapting the transmission rate accordingly \cite{dai2022nonlinear,wang2023wireless}. This end-to-end rate-distortion (RD) optimized JSCC framework, particularly in scenarios involving high-resolution images and videos, has shown significant system coding gains over basic DeepJSCC approaches. Its latest version \cite{wang2023improved} has exhibited on-par or superior end-to-end rate-distortion performance compared to the state-of-the-art engineered source and channel codecs, such as VTM (VVC intra coding) + 5G LDPC.

Optimizing traditional rate-distortion metrics, i.e., minimizing the expected data transmission rate under a distortion constraint such as MSE, does not yet ideally cater to human perceptual qualities, often at the expense of realism. Here, realism is mathematically connected to the statistical fidelity or $f$-divergence between the distributions of the reconstructed and original images. This fidelity is typically optimized using an adversarial discriminator loss \cite{mentzer2020high} and assessed using metrics like the Fréchet Inception Distance (FID) \cite{heusel2017gans}. Intuitively, rate-distortion (RD) optimization tends to yield pixel-wise averages of plausible solutions that are overly smooth and misaligned with human perceptual interpretation \cite{ledig2017photo}. Theoretical and empirical studies have further revealed that achieving perfect realism may result in at most a two-fold increase in optimal MSE, which yet significantly impacts the perceived quality of reconstructed images \cite{blau2018perception,blau2019rethinking}.

These findings have catalyzed the development of codecs that prioritize human perceptual quality, referred to as ``generative compression'' or ``generative transmission'' codecs. Such codecs aim for reconstructions that, while potentially differing from the original, are indistinguishable to observers, thereby focusing on end-to-end rate-distortion-perception (RDP) optimization. This approach has gained traction in both neural compression \cite{mentzer2020high, mentzer2022neural, yang2024lossy, muckley2023improving} and end-to-end neural transmission strategies \cite{wang2022perceptual, dai2022nonlinear, yue2023learned}. 

\subsection{Motivation}

{
Despite the effectiveness of existing end-to-end transmission approaches, we identify two significant limitations as follows:}
\begin{itemize}
	\item \emph{Inferior fidelity:} Common artifacts such as blurring and aliasing are particularly noticeable under challenging conditions including low transmission rates and poor wireless channel quality. These artifacts lead to unrealistic and inconsistent reconstructions, despite the incorporation of perceptual or adversarial losses during training. 
	{
	Compared with separation-based counterparts (e.g., MS-ILLM \cite{muckley2023improving} + 5G LDPC), existing end-to-end optimized JSCC codecs \cite{wang2022perceptual} still lag behind in RDP performance by a substantial margin. Following recent theoretic analysis on the perception-robustness trade-off \cite{ohayon2023reasons, ohayon2023perception}, we suggest that a primary reason for the RDP performance gap stems from the deterministic decoding paradigm. Generative transmission codecs \cite{wang2022perceptual} are primarily optimized to be robust against stochastic degradations introduced by the wireless channel, which profoundly hinder their perceptual quality~\cite{ohayon2023perception}.
	
	
	\item \emph{Inferior generalization and robustness:} 
	Channel characteristics are implicitly learned during the end-to-end codec training process. When the channel-received signal often experiences severe wireless-related degradations that were not encountered during training, these systems often underperform or even fail to produce meaningful reconstructions. In scenarios where retransmission is not permitted, a feasible solution is to generate numerous data samples from the natural data manifold and select the sample that best matches the corrupted signal as the decoding result. However, existing autoencoder-based transmission systems can only reconstruct input data, without the ability to generate new data, thereby missing the last opportunity to enhance robustness.
}
\end{itemize}
%
%
%
%

In this paper, we aim to integrate additional prior knowledge and leverage the superior capabilities of \emph{stochastic sampling} methods over deterministic approaches to develop a \emph{generative end-to-end communication system}.
Our design objectives are articulated as follows:
\begin{itemize}
	\item \emph{Stochastic posterior sampler as the decoder:} 
	{ To simultaneously achieve perfect perceptual quality and maintain consistency with the ground truth over dynamic wireless channels, the decoding algorithm must inherently operate as a \emph{posterior sampler} and, consequently, must be stochastic in nature \cite{ohayon2023reasons, ohayon2023perception}.}
	
	\item \emph{Utilizing channel received signal to guide posterior sampling:} Unlike coarse-grained conditions used in generation such as text or sketch descriptions, in communication setup, the channel received signal provides a natural, fine-grained condition for guiding posterior sampling. This controllable generative process allows decoder to produce reliable and consistent results with the ground truth.
	
\end{itemize}

{ To achieve this, we take inspiration from existing research on inverse problems, which seeks to infer the unknown source $\bm{x}$ from indirect and noisy observations $\bm{y}$. Recently, inverse problem solvers have incorporated diffusion models \cite{song2019generative, song2020score, dhariwal2021diffusion} as plug-and-play priors, establishing themselves as the \emph{de facto} solution in natural image restoration tasks \cite{chung2022improving, chung2022diffusion, wang2022zero, zhu2023denoising}. 
The core idea is to modify the diffusion trajectory by incorporating a likelihood term to generate samples from the posterior distribution $p(\bm{x}|\bm{y})$, which under Bayes rule can be expressed as $p(\bm{x}|\bm{y}) \propto p(\bm{x})p(\bm{y}|\bm{x})$.
Here, the likelihood term, $p(\bm{y}|\bm{x}_t)$, at an intermediately reverse diffusion timestep $t$, is generally intractable. This intractability has prompted the development of various approximations \cite{song2021solving, chung2022diffusion, song2023pseudoinverse, pmlr-v202-song23k}.
Most diffusion inverse solvers are designed to handle simple forward operators (e.g., down-sampling and motion blurring) and operate under conditions of weak measurement noise (high signal-to-noise ratio, typically $>$10dB). However, these approaches may prove ineffective in end-to-end communication scenarios, where the channel-received signal is generated by highly nonlinear neural networks with tens of millions of parameters and is subjected to severe channel fading and noise.
}
	

\subsection{Contribution}

With the aforementioned motivations, in this paper, we propose \emph{DiffCom}, a novel end-to-end communication paradigm that exploits pre-trained diffusion generative priors while leveraging stochastic generative components. Unlike conventional methods that utilize deterministic decoders, \emph{DiffCom} employs a unified source-channel encoder but decodes the channel received signal using probabilistic generative models. 

For an image source, we seek an image $\bm{x}$ that resides on the generative prior. This image, after being processed by the source-channel encoder function $\mathcal{E}(\cdot)$ and the wireless channel operator $\mathcal{W}(\cdot)$, aligns closely with the corrupted channel received signal $\bm{y}$ in the latent space. 
To achieve this, \emph{DiffCom} leverages diffusion models' strong generative priors to reverse the corrupted channel received signal $\bm{y}$.
Specifically, we integrates off-the-shelf score-based log-likelihood gradient steps towards the received signal ($\nabla_{\bm{x}} \log p(\bm{y} | \bm{x})$) into the intermediate ancestral sampling stages of a pre-trained diffusion model with generative prior ($\nabla_{\bm{x}} \log p(\bm{x})$). This integration essentially employing the channel received signal as a fine-grained condition to guide diffusion posterior sampling.
In this manner, \emph{DiffCom} always pursues realistic image reconstructions, whose consistency with the original image reflects the uncertainties inherent in the received signal. 
Moreover, this approach is not restricted to specific types of wireless degradation.
By incorporating these factors into the construction of the channel operator $\mathcal{W}(\cdot)$, \emph{DiffCom} exhibits robustness against a wide range of unforeseen wireless transmission impairments, including mismatched channel signal-to-noise ratio (CSNR), previously unseen fading conditions, inaccurate or missing channel estimations, peak-to-average power ratio (PAPR) reduction, inter-symbol interference, etc.

{ However, the inherent stochasticity of the generative process poses significant challenges in effectively and precisely guiding the generator using the channel-received signal. 
Compared to most diffusion inverse solvers \cite{chung2022improving, chung2022diffusion, wang2022zero, zhu2023denoising} on image restoration, we deals with much more noisier measurement signal in \emph{latent space} and operates inversion through a \emph{highly nonlinear neural network}.
As a result, for vanilla \emph{DiffCom}, we find when a small-enough measurement distance has been achieved, the resulting reconstruction is realistic enough but possibly not faithfully represent the original data. 
This ambiguity, driven by high-power noise and the nonlinear nature of the neural networks, considerably increases the difficulty of achieving stable and efficient diffusion posterior sampling, thereby impacting the quality of the finally decoded images.}

To address this issue, we develop \emph{HiFi-DiffCom} to attain stable and efficient diffusion posterior sampling. 
We propose penalizing posterior mean-based latent samples by further enforcing consistency in the source space and tailoring the sampling process to the quality of measurements. 
Through these methods, \emph{DiffCom} demonstrates the ability to stably generate realistic images that also retain faithful details consistent with the ground truth. 
Additionally, to enhance the robustness of \emph{DiffCom} against unexpected wireless transmission degradations, we further extend the system to include blind posterior sampling capabilities. 
This is achieved by constructing an additional diffusion prior on the channel response distribution, enabling \emph{DiffCom} to adapt to different transmission scenarios effectively. 
In summary, our \emph{DiffCom} series, i.e., the channel received signal guided diffusion posterior sampling framework, not only addresses the limitations of traditional deterministic decoders but also sets a new benchmark in designing reliable and controllable generative communication systems, while remaining robustness and generalization superiority.

Our core contributions in this paper are threefold:
\begin{enumerate}
	\item \emph{DiffCom Framework:} We introduce \emph{DiffCom}, a novel generative end-to-end communication paradigm that utilizes the corrupted received signal as a natural condition to guide pre-trained unconditional diffusion generators. \emph{DiffCom} sets a new benchmark in end-to-end transmission by achieving state-of-the-art performance across multiple perceptual quality metrics and demonstrating robustness against diverse wireless-related degradations.
	
	\item \emph{Enhanced Posterior Sampling Efficiency:} Leveraging the identical received signal as RD-optimized methods, \emph{DiffCom} is capable of simultaneously decoding with optimal perceptual quality and minimal MSE. We enhance this process by introducing an extra confirming constraint that ensures generated samples remain on the manifold of real data, and by accelerating the sampling process to adapt to the quality of the received signal. These innovations have proven to enhance the sampling efficiency under various channel conditions in \emph{DiffCom}.
	
	\item \emph{Extension to Blind Cases:} We further extend \emph{DiffCom} by constructing a diffusion prior on the channel response distribution, enabling the system to tackle extreme blind transmission scenarios. This extension allows for the joint estimation of channel parameters and source data in a coarse-to-fine manner, leveraging the interplay between source and channel diffusion models.
\end{enumerate}

We have also noted other attempts to integrate diffusion models into end-to-end wireless visual communication systems \cite{wu2024wireless, yilmaz2023high}. These approaches primarily utilize diffusion models either as denoisers for pre-processing the received latents, requiring additional joint training, or as post-processing modules to refine degraded reconstructions. Such strategies have resulted in pronounced blurring artifacts in the visual outcomes. {Conceptually, from the perspective of data-processing inequality in Shannon theory \cite{cover1999elements}, both extra pre-processing and post-processing operations may lead to some additional information loss, failing to accurately represent the input or exacerbating unwanted artifacts}. In contrast, our \emph{DiffCom} leverages the raw channel received signal to guide posterior generative sampling directly. This approach enables the production of realistic visual results with details that are faithfully consistent with the ground truth, effectively avoiding the pitfalls of separate processing stages.

\section{Problem Formulation}

Let us assume an input data sample $\bm{x}$, represented as a vector of intensities $\bm{x} \in \mathbb{R}^m$, whose distribution $p(\bm{x})$ could be modeling, for example, audio signals, natural images or videos. 
Under the JSCC setup, the goal of end-to-end data transmission is to create an encoder function $\mathcal{E}$, encoding $\bm{x}$ to a vector of complex-valued channel-input symbols $\bm{z} \in \mathbb{C}^k$, and a decoder function $\mathcal{D}$, recovering the corrupted received signal $\bm{y} = \mathcal{W}_{\bm{h}^*}(\bm{z}) + \bm{n}$ back to the input sample.
We also follow the setup of orthogonal frequency division multiplexing (OFDM) transmission \cite{yang2022ofdm}, where the function $\mathcal{W}_{\bm{h}^*}(\cdot)$ encapsulates transmission-related signal processing operations and wireless-related degradations, such as add pilots, add cyclic-prefix (CP), PAPR reduction, channel fading with channel impulse response $\bm{h}^* \in \mathbb{C}^L$ with $L$ multipaths ($\bm{h}^*$ denotes the ground truth channel response), and so on.
Addictive noise $\bm{n}$ is independently sampled from Gaussian distribution $\bm{n} \sim \mathcal{CN}(0, \sigma_n^2 \bm{I}_k)$ with noise power $\sigma_n^2$.

\subsection{Rate-Distortion Optimized End-to-End Transmission}

Typically, $\mathcal{E}$ and $\mathcal{D}$ are jointly optimized to minimize the RD objective in the presence of wireless degradations, formulated as a linear combination of a bandwidth rate and an end-to-end distortion term:
\begin{equation}
	\mathbb{E}_{\bm{x} \sim p(\bm{x})} \mathbb{E}_{\bm{h}^* \sim p(\bm{h}^*)} \left[
	\mathcal{L}_R(\bm{z}) + \lambda \mathcal{L}_D\left(\mathcal{D}\left(\mathcal{W}^{-1}_{\bm{h}}(\bm{y})\right), \bm{x}\right) \right],
\end{equation}
where $\mathcal{L}_R$ is the bandwidth rate term measured by \emph{channel bandwidth ratio} (CBR) $\rho = k / m$, $\mathcal{L}_D(\cdot, \cdot)$ denotes a distance measure typically formulated by signal-level distortions such as mean-squared error (MSE), and $\mathcal{W}_{\bm{h}}^{-1}$ indicates the necessary explicit inverse operations on~$\bm{y}$ with estimated channel response ${\bm{h}}$, such as channel estimation, and equalization. 

For simplicity, in fixed-length coding methods \cite{djscc, xu2021wireless, yang2022ofdm, yang2024swinjscc}, the bandwidth rate term $\mathcal{L}_R$ is a constant value specified to the autoencoder bottleneck dimension, in which case $\mathcal{L}_R$ can be dropped from the optimization formulation. 
In contrast, for variable-length JSCC codecs, $\mathcal{L}_R$ is instance-adaptive, which can be optimized through learned entropy estimation \cite{dai2022nonlinear, wang2023wireless, wang2023improved}, policy networks with categorical reparameterization \cite{yang2022deep}, and reinforcement learning approaches \cite{tung2022deepwive}.

In general, despite the natural uncertainty introduced by wireless channel, both $\mathcal{E}$ and $\mathcal{D}$ are deterministic for the majority of end-to-end learned transmission methods. 
In this case, we empirically find that as CBR $\rho$ or CSNR decreases, their reconstructions will incur increasing degradations, such as blurring, blocking, and aliasing artifacts. 
The inherent reason for this discrepancy is that low signal-level distortions do not necessarily equate to high perceptual quality. 
The latter is more accurately described as a divergence between the distributions of the source and the reconstructed data, with perfect perceptual quality achieved when the two distributions are identical \cite{blau2019rethinking, blau2018perception}.

We have also observed that these end-to-end trained methods often perform worse than expected or even fail to function properly when evaluated under unseen wireless environments, such as mismatched CSNR and unseen fading. 
The characteristics of wireless-related degradations are only implicitly considered during the end-to-end codec training, leading to inferior generalization when the testing channel response distribution differs from the one used during model training. 
In such cases, deterministic decoders typically require either additional training \cite{yang2022ofdm} or online adaptation \cite{dai2023toward} to accommodate new channel domain knowledge.

\begin{figure}[t]
	\setlength{\abovecaptionskip}{0cm}
	\setlength{\belowcaptionskip}{0.cm}
	\centering
	\includegraphics[width=\linewidth]{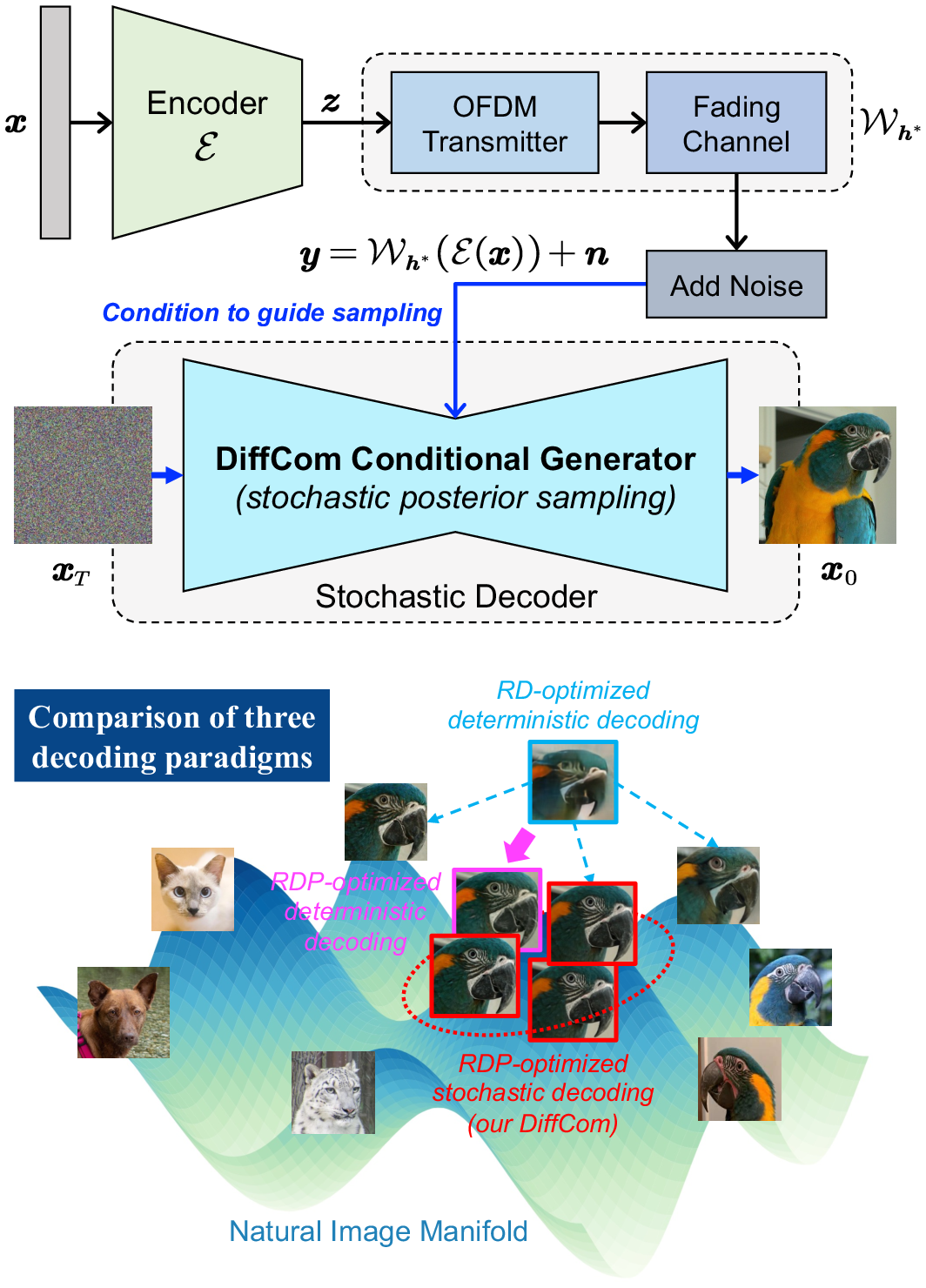}
	\caption{Overview of our \emph{DiffCom} system architecture, and the illustration of three decoding paradigms. (1) The RD-optimized deterministic decoding result appears overly smooth reconstructions due to the pixel-wise averaging of potential solutions in the source space; (2) The RDP-optimized deterministic decoding promotes reconstructions towards regions of the search space that are likely to contain photo-realistic images, thereby aligning more closely with the natural image manifold; (3) Our RDP-optimized stochastic decoding employs diffusion posterior sampling to generate a diverse array of solutions. The channel received signal serves as a fine-grained condition to guide the sampling process, consistently yielding high-fidelity outcomes. This paradigm enhances both the robustness and generalization capabilities of the decoding process.}
	\label{fig:framework}
\end{figure}


\subsection{Distribution-Preserving Generative Transmission}
To address above limitations of RD optimized end-to-end transmission paradigm, we take further insights from rate-distortion-perception theory \cite{blau2018perception}, which suggest that perfect realism can be achieved with no more than a twofold increase in distortion from the RD optimal codec. 
This insight has catalyzed the development of advanced generative compression codecs \cite{tschannen2018deep, mentzer2020high, mentzer2022neural}. 
Motivated by these works, in this paper, we propose a novel generative transmission paradigm ``\emph{DiffCom}'' as shown in Fig. \ref{fig:framework}, which primarily focusing on minimizing the divergence between the distributions of source and reconstruction, while searching for samples that consistent with the received signal at the same time. 
In this manner, our transmission paradigm strives to preserve the original distribution, ensuring that reconstructions are realistic and free of artifacts and unreasonable results even in case of unseen wireless environments.

To achieve this, we leverage the RD-optimized deterministic encoder $\mathcal{E}$ but decodes stochastically with a pre-trained unconditional generator.
To be specific, we aim to find an image $\bm{x}_0$ that lives on the generative prior $p_{{\bm \theta}}(\bm{x}_0)$ and, after being processed through $\mathcal{E}$ and $\mathcal{W}_{\bm{h}}$, closely aligns with the corrupted received signal (measurement) $\bm{y}$ in latent space.
Formally, the problem we aim to solve at the receiver is expressed as:
\begin{equation}
	\label{eq:dis}
	\argmin_{\bm{x}_0}{\|\bm{y} - \mathcal{W}_{\bm{h}}(\mathcal{E}(\bm{x}_0))\|_2^2} ,~\textrm{s.t.}~ \bm{x}_0 \sim p_{{\bm \theta}}(\bm{x}_0),
\end{equation}
where $\mathcal{W}_{\bm{h}}(\cdot)$ represents the wireless transmission-related operations and degradations, which are simulated by the receiver using the estimated parameter $\bm{h}$.
Our proposed generative transmission framework is easy for implementation, since we can employ off-the-shelf JSCC encoder $\mathcal{E}$ and a pre-trained unconditional generator.
Furthermore, our framework naturally exhibits superior robustness to unexpected perturbations in $\mathcal{W}_{\bm{h}^*}$, offering a significant advantage over existing deterministic decoders that fragile to unseen input signal distribution.

We note that the problem defined in \eqref{eq:dis} has a similar form with the generative model inversion works \cite{chung2022improving, chung2022diffusion, zhu2023denoising}, which employ generative priors to solve linear ill-posed problems with known degradations in \emph{source pixel space}, such as super-resolution, inpainting, and other restoration tasks. 
However, our \emph{DiffCom} still differs clearly from them as inverting measurement signal \emph{in latent space} and through a \emph{highly-nonlinear} neural encoder.
Moreover, in the context of wireless transmission, \emph{DiffCom} deals with much more noisy measurement signal, and the degradations estimated by receiver $\mathcal{W}_{\bm{h}}$ can also be different from the ground truth~$\mathcal{W}_{\bm{h}^*}$, which is typically intractable. These factors significantly elevate the difficulty of stable and efficient generative sampling, subsequently affecting the quality of the decoded images. 


\section{Methodology}\label{section_method}

Due to the stochastic nature of the generative process, it is hard to solve \eqref{eq:dis} under strict distribution constraint. 
Inspired by restoration tasks, such as deep image prior \cite{ulyanov2018deep}, we relax this problem by optimizing the following objective function
\begin{equation}
	\label{eq:hqs}
	\argmin_{\bm{x}_0}{\underset{\text{for realism}}{\underbrace{\mathcal{P}(\bm{x}_0)}} + \zeta \cdot \underset{\text{for latent consistency}}{\underbrace{\|\bm{y} - \mathcal{W}_{\bm{h}}(\mathcal{E}(\bm{x}_0))\|_2^2}}},
\end{equation}
where $\mathcal{P}(\cdot)$ represents a regularization function of the generative prior, such as $-\log p_{{\bm \theta}}(\bm{x}_0)$, and $\zeta$ controls the \emph{guidance strength} of the channel received signal.
The first term of \eqref{eq:hqs} is responsible for \emph{realism term}, which encourages $\bm{x}_0$ to live close to the generative prior.
Whereas the second term is the \emph{latent consistency term}, which quantifies the latent distance between the received corrupted feature $\bm{y}$ and the feature extracted from the generated image~$\bm{x}_0$. 

The overall performance of this generative end-to-end transmission framework depends on three key factors:
\begin{enumerate}
	\item How much information $\bm{y}$ and $\bm{h}$ preserve about $\bm{x}$ and the ground truth channel response $\bm{h}^*$, which is primarily dependent on the performance of the JSCC codec and the channel estimation algorithm.
	
	\item How accurately the prior-imposing function $\mathcal{P}(\cdot)$ estimates the true prior, which relies on the distribution coverage capability of the adopted generative models.
	
	\item How effectively the optimization procedure identifies the minima, dependent on the efficiency of sampling algorithms used in conjunction with the specific generator.
\end{enumerate}
The first factor has been extensive investigated by existing end-to-end transmission methods and wireless channel estimation algorithms.
Therefore, this paper primarily focuses on the last two factors.



\subsection{Unconditional Diffusion Probabilistic Models}

\emph{DiffCom} employs diffusion probabilistic models \cite{song2020score} as a powerful generative prior to stochastically decode channel-received signals. As a preliminary, we offer a brief overview of diffusion models.
Diffusion models generate images with forward process and reverse process defined over $T$ reversible time steps. The forward process is to diffuse the data distribution $p(\bm{x})$ to a standard Gaussian distribution, $p_{T}(\bm{x}) = \mathcal{N}(\bm{0}, \bm{I})$, by injecting noise continuously at each step $t$. Under the variance-preserving (VP) setting \cite{ho2020denoising}, the forward \emph{noising transition kernel} is parameterized as
\begin{equation}
	q(\bm{x}_t|\bm{x}_{t-1}) =
	\mathcal{N}(\bm{x}_t;\sqrt{1-\beta_t}\bm{x}_{t-1}, \beta_t\bm{I}_m),
	\label{eq:diffused_kernel}
\end{equation}
where $\{\beta_t\}_{t=1}^T$ denotes a pre-defined or learned noise variance schedule. 
With reparametrization trick, the marginal distribution at arbitrary timestep $t$ can be computed analytically:
\begin{equation}
	q(\bm{x}_t|\bm{x}_0) =
	\mathcal{N}(\bm{x}_t;\sqrt{\overline{\alpha}_t}\bm{x}_0, (1-\overline{\alpha}_t)\bm{I}_m),
	\label{eq:xt_cond_x0}
\end{equation}
where $\alpha_t = 1-\beta_t$ and $\overline{\alpha}_t = \prod_{i=0}^{t} \alpha_t$ fix the noise strength, varying from $\overline{\alpha}_0  \approx 1$ for no noise to $\overline{\alpha}_T  \approx 0$ for pure Gaussian noise. 

Under the continues time viewpoint, the forward diffusion process can also be written as a stochastic differential equation (SDE) of the form \cite{song2020score}:
\begin{equation}
	d\bm{x} = -\frac{\beta_t}{2}\bm{x}_t dt + \sqrt{\beta_t}d{\bm{w}},
	\label{eq:reverse_sde}
\end{equation}
where $\beta_t > 0$ controls the speed of diffusion and $\bm{w}$ follows the standard Wiener process. To sample an image, we can employ the reverse SDE, which takes the form:
\begin{equation}
	d\bm{x} = \left[-\frac{\beta_t}{2}\bm{x}_t - \beta_t\nabla_{\bm{x}_t} \log p({\bm{x}_t})\right]dt + \sqrt{\beta_t}d\bar{\bm{w}},
	\label{eq:reverse_sde}
\end{equation}
where $\bar{\bm{w}}$ is the standard Wiener process running backward in time. 
The quantity $\nabla_{\bm{x}_t} \log p({\bm{x}_t})$ is the \emph{prior score}, which can be practically estimated using a neural network $s_{\bm \theta}: \mathbb{R}^m \times [0, 1] \rightarrow \mathbb{R}^m$ with the U-net architecture parameterized by ${\bm \theta}$. The training objective of score function $s_{\bm \theta}$ is denoising score matching~\cite{song2020score}: 
\begin{equation}
	\argmin_{\bm \theta} \mathbb{E}_{t, \bm{x}_0, \bm{x}_t|\bm{x}_0} \left[\|s_{\bm \theta}(\bm{x}_t, t) - \nabla_{\bm{x}_t}\log p(\bm{x}_t|\bm{x}_0)\|_2^2\right],
	\label{eq:dsm}
\end{equation}
where $t \sim \mathcal{U}(0, T)$, $\bm{x}_0 \sim p(\bm{x})$, and $\bm{x}_t|\bm{x}_0$ is sampled with \eqref{eq:xt_cond_x0}.
In the discretized form, the forward and reverse SDE is formulated as
\begin{equation}
	\bm{x}_t = \sqrt{1 - \beta_t} \bm{x}_{t-1} + \sqrt{\beta_t} \bm{\epsilon}_{t-1}, 
	\label{eq:forward_step}
\end{equation}
\begin{equation}
	\bm{x}_{t-1} = \frac{1}{\sqrt{1 - \beta_t}} (\bm{x}_{t} + \beta_t \nabla_{\bm{x}_t} \log p({\bm{x}_t})) + \sqrt{\beta_t} \bm{\epsilon}_{t}, 
	\label{eq:reverse_step}
\end{equation}
respectively, where $\bm{\epsilon} \sim \mathcal{N}(0, \bm{I}_m)$.

\subsection{The Basic Architecture of DiffCom}

With score-based diffusion generative modeling, we are able to generate samples in the unconditional manner from the prior data distribution by utilizing the pre-trained score function $s_{\bm \theta}(\bm{x}, t)$. 
In the context of end-to-end transmission, however, our goal shifts to sampling from the posterior distribution $p(\bm{x} | \bm{y})$, which requires carefully conditioning the stochastic generative process on the corrupted channel received signal.
Therefore, we modify \eqref{eq:reverse_sde} by solving the conditional reverse SDE to facilitate the posterior sampling:
\begin{equation}
	d\bm{x} = \left[-\frac{\beta_t}{2}\bm{x} - \beta_t\nabla_{\bm{x}_t} \log p({\bm{x}_t} | \bm{y})\right]dt + \sqrt{\beta_t}d\bar{\bm{w}}.
	\label{eq:reverse_conditional_sde}
\end{equation}
Herein, the \emph{posterior score} $\nabla_{\bm{x}_t} \log p({\bm{x}_t} | \bm{y})$ is intractable to compute.
A straightforward approach, as suggested in \cite{rombach2022high}, is to train a new conditional diffusion model $s_{\bm \theta}(\bm{x}, \bm{y}, t)$ explicitly with paired data points $\{(\bm{x}_i,\bm{y}_i)\}_{i=1}^N$. 
However, such a supervised training method will be tightly restricted to the specific operations and channel response distributions encapsulated by $\mathcal{W}_{\bm{h^*}}$ in our communication setup, potentially preventing models from generalizing effectively under varying wireless conditions.
Therefore, we explore the use of diffusion inverse solvers to infer the lost information due to nonlinear encoding and channel corruptions, which have shown impressive results in vanilla  image restoration tasks \cite{song2021solving, chung2022improving, chung2022diffusion, zhu2023denoising}.

In specific, the posterior score can be decomposed through Bayesian rule:
\begin{equation}
	\nabla_{\bm{x}_t} \log p(\bm{x}_t|\bm{y}) = \nabla_{\bm{x}_t} \log p(\bm{x}_t) + \nabla_{\bm{x}_t} \log p(\bm{y}|\bm{x}_t),	\label{eq:bayes}
\end{equation}
where the prior score can be approximated with the pre-trained diffusion model, i.e., $\nabla_{\bm{x}_t} \log p(\bm{x}_t) \approx s_{\bm \theta}(\bm{x}_t, t)$.

\begin{figure}[t]
	\setlength{\abovecaptionskip}{-0cm}
	\setlength{\belowcaptionskip}{0cm}
	\begin{center}
		\begin{tikzpicture}[
			Nodes/.style={circle, draw=gray!80, fill=gray!10, very thick, minimum size=10mm},
			Nodex/.style={circle, draw=red!80, fill=red!5, very thick, minimum size=10mm},
			Nodey/.style={circle, draw=blue!80, fill=blue!5, very thick, minimum size=10mm},
			Noded/.style={circle, draw=green!80, fill=green!10, very thick, minimum size=10mm}
			]
			\node[Nodex]    (x0)   {${\bm x}_0$};
			\node[Nodey]  (y) [below left = of x0]  {$\bm y$};
			\node[Nodex]  (xt)   [below right= of x0] {${\bm x}_t$};
			\node[Nodes]  (x) [left = 2.5of y]{$\bm x$} ;

			%
			%
			\draw [->, thick] (x) to [out=0,in=180] node[above] {$p({\bm{y}}|{\bm{x}})$} node[below] {$\mathcal{W}_{\bm{h}^*}(\mathcal{E}(\bm{x})) + \bm{n}$}  (y);
			\draw [->, thick] (x0) to [out=-150,in=60] node[above, sloped] {$p({\bm{y}}|{\bm{x}_0})$}  (y);
			\draw [->, thick] (x0) to [out=-30,in=120] node[above, sloped] {$q({\bm x}_t|{\bm x}_0)$} (xt);
			\draw [->, thick,dotted] (xt) to [out=150,in=-60] node[below, sloped] {$p({\bm x}_0|{\bm x}_t)$} (x0);
			
		\end{tikzpicture}
		\caption{Probabilistic graph of standard \emph{DiffCom}. Solid line: tractable, dotted line: intractable in general.}
		\label{fig:prob_graph_diffcom}
	\end{center}
\end{figure}
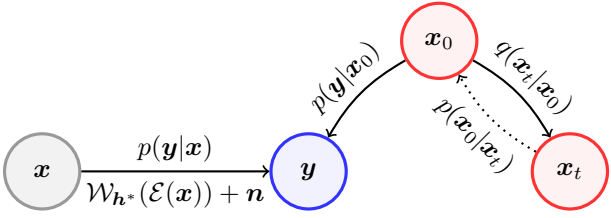

For the time-conditional \emph{likelihood score} $\nabla_{\bm{x}_t} \log p(\bm{y}|\bm{x}_t)$, as shown in Fig. \ref{fig:prob_graph_diffcom}, there does not exist explicit dependency between $\bm{y}$ and $\bm{x}_t$.
What we can exploit is the measurement model $\bm{x}_0 \rightarrow \bm{y}$ with estimated channel response $\bm{h}$, which is represented as
\begin{align}
	p(\bm{y} | \bm{x}_0) &= \mathcal{N}(\bm{y}; \mathcal{W}_{\bm{h}^*}(\mathcal{E}(\bm{x}_0)), \sigma_n^2 \bm{I}_k), \notag \\
	&\approx \mathcal{N}(\bm{y}; \mathcal{W}_{\bm{h}}(\mathcal{E}(\bm{x}_0)), \sigma_n^2 \bm{I}_k).
	\label{eq:measurement_model}	
\end{align}
To seek for a tractable surrogate, leveraging the idea of diffusion posterior sampling (DPS) \cite{chung2022diffusion}, the likelihood score term can be approximated by
\begin{align}
	\nabla_{\bm{x}_t} \log p(\bm{y}|\bm{x}_t) &= \nabla_{\bm{x}_t} \log \mathbb{E}_{\bm{x}_0 \sim p(\bm{x}_0 | \bm{x}_t)}[p(\bm{y}|\bm{x}_0)], \notag \\
	&\overset{(a)}{\approx} \nabla_{\bm{x}_t} \mathbb{E}_{\bm{x}_0 \sim p(\bm{x}_0 | \bm{x}_t)}[\log p(\bm{y}|\bm{x}_0)], \notag \\
	&\overset{(b)}{\approx} \nabla_{\bm{x}_t} \log p(\bm{y}|\mathbb{E}_{\bm{x}_0 \sim p(\bm{x}_0 | \bm{x}_t)}[\bm{x}_0]),
	\label{eq:app_dps}
\end{align}
where the two approximations follow Jenson inequity. \cite{song2021solving} has pointed out that there are indeed two biased approximations in \eqref{eq:app_dps}:
\begin{itemize}
	\item (a) is always biased as $\log(\cdot)$ is concave.
	\item (b) is also biased as the posterior sample ${\bm{x}}_{0|t}$ is different from the mean $\hat{\bm{x}}_{0|t} := \mathbb{E}_{\bm{x}_0 \sim p(\bm{x}_0 | \bm{x}_t)}[\bm{x}_0]$.
\end{itemize}
The posterior mean $\hat{\bm{x}}_{0|t} := \mathbb{E}_{\bm{x}_0 \sim p(\bm{x}_0 | \bm{x}_t)}[\bm{x}_0]$ can be obtained through the Tweedie's formula
\begin{align}
	\hat{\bm{x}}_{0|t} &= \frac{1}{\sqrt{\overline{\alpha}_{t}}}\left( \bm{x}_{t} + \left({1-\overline{\alpha}_{t}}\right) \nabla_{\bm{x}_t} \log p(\bm{x}_t) \right), \notag \\
	&\approx \frac{1}{\sqrt{\overline{\alpha}_{t}}}\left( \bm{x}_{t} + \left({1-\overline{\alpha}_{t}}\right) s_{\bm \theta}(\bm{x}_{t},t) \right).
	\label{eq:posterior_mean}
\end{align}
Substituting \eqref{eq:measurement_model} and \eqref{eq:posterior_mean} into \eqref{eq:app_dps}, the likelihood score can be concluded as
\begin{align}
	\nabla_{\bm{x}_t} \log p(\bm{y}|\bm{x}_t) \approx - \frac{1}{\sigma_n^2} \nabla_{\bm{x}_t} \underset{\text{measurement distance:~}\mathcal{L}_m}{\underbrace{\|\bm{y} - \mathcal{W}_{\bm{h}}(\mathcal{E}(\hat{\bm{x}}_{0|t}))\|_2^2}}.
	\label{eq:score_likelihood}
\end{align}
As a result, we can decode $\bm{y}$ stochastically with the posterior score
\begin{equation}
	\nabla_{\bm{x}_t} \log p(\bm{x}_t|\bm{y}) \approx s_{\bm \theta}(\bm{x}_t, t) - {\zeta_t} \nabla_{\bm{x}_t}{\|\bm{y} - \mathcal{W}_{\bm{h}}(\mathcal{E}(\hat{\bm{x}}_{0|t}))\|_2^2},	\label{eq:score_posterior}
\end{equation}
where $\zeta_t$ controls the guidance strength of time step $t$. 

Recall the optimization constraint in \eqref{eq:hqs}, now we tackle it iteratively in a coarse-to-fine manner.
In practice, for each time step $t$, our method alternates between two operations:
\begin{itemize}
	\item \emph{Enforce realism}: We sample the approximate posterior mean $\hat{\bm{x}}_{0|t}$ using an unconditional diffusion reverse step as \eqref{eq:posterior_mean}, which aims to align the sampled data closer to the generative prior, reinforcing the realism of produced image.
	\item \emph{Enforce data consistency}: We perform gradient descents with $\hat{\bm{x}}_{0|t}$ to update the ancestral sample $\bm{x}_{t-1}$. This step gradually refines samples to be consistent with the latent measurement (channel received signal) $\bm{y}$.
\end{itemize}
Along the iteration progresses, each step progressively refines the solution, converging toward the optimization goal outlined in \eqref{eq:hqs}. 
A promising solution that meets our optimization objectives is achieved when $t=0$.
The specific steps of standard \emph{DiffCom} posterior sampling process are detailed in Algorithm~\ref{alg1}.
 
{ As a natural extension of the DPS \cite{chung2022diffusion} idea to end-to-end communications, \emph{DiffCom} performs the entire inversion process within the latent space. This strategy allows \emph{DiffCom} to approximate the unknown clean latent signal $\bm{y}_0$ with the encoded version of the conditional expectation of the clean image $\bm{x}_0$, without involving the deterministic decoder $\mathcal{D}$.
We emphasize that this method is applicable for majority RD-optimized JSCC codecs, requiring only their encoder components and a pre-trained unconditional diffusion model.
Moreover, leveraging the received signal in the latent space provides superior perceptual quality and robustness, compared to performing restoration after passing through the decoder, thereby avoiding additional information loss, which will be detailed in the ablation study.
}

\begin{figure}
	\vspace{-0.5em}
\begin{algorithm}[H]
	\small
	\caption{Standard Posterior Sampling of DiffCom}\label{alg1}
	\begin{algorithmic}[1] 
		\Require {$\bm{y}$, $\mathcal{E}$, $\mathcal{W}_{\bm{h}}$, $T$, $\{\zeta_t\}^T_{t=1}$, $\{\sigma_t\}^T_{t=1}$, $s_{\bm \theta}(\cdot,\cdot)$.}
		\State Initialize $\bm{x}_T \sim \mathcal{N} (\bm{0}, \bm{I}_m)$ 
		\State $ \bm{h} = \emph{Channel-Estimation}(\bm{y}_{\text{pilot}})$
		\For {$t=T, \ldots , 1$}
		\LineComment{Estimate posterior mean $\mathbb{E}[\bm{x}_0|\bm{x}_t]$ }
		\State $\hat{\bm{s}} \gets s_{\bm \theta}(\bm{x}_{t},t)$
		\State $\hat{\bm{x}}_{0|t} \gets \frac{1}{\sqrt{\overline{\alpha}_{t}}}\left( \bm{x}_{t} + \left({1-\overline{\alpha}_{t}}\right) \hat{s} \right)$
		\LineComment{Sample \emph{i.i.d.} Gaussian noise}
		\State $\bm{\epsilon} \sim \mathcal{N} (\bm{0}, \bm{I}_m)$ \textbf{if} $t > 0$, \textbf{else} $\bm{\epsilon}=\bm{0}$
		\LineComment{Diffusion ancestral sampling}
		\State $\bm{x}_{t-1}^\prime \gets \frac{\sqrt{\alpha_{t}} (1 - \overline{\alpha}_{t-1})}{1 - \overline{\alpha}_{t}}\bm{x}_{t} + \frac{\sqrt{\overline{\alpha}_{t-1}} {\beta}_{t}}{1 - \overline{\alpha}_{t}}\hat{\bm{x}}_{0|t} + \tilde{\sigma}_t \bm{\epsilon}$ 
		\LineComment{Enforcing data consistency in latent space}
		\State $\mathcal{L}_m \gets \Vert\bm{y} -\mathcal{W}_{\bm{h}}\left(\mathcal{E}\left(\hat{\bm{x}}_{0|t}\right)\right)\Vert_2^2$
		\LineComment{Finish one step posterior sampling}
		\State $\bm{x}_{t-1} \gets \bm{x}_{t-1}^\prime - \zeta_t \nabla_{\bm{x}_t} \mathcal{L}$
		\EndFor
		\State \textbf{return} $\bm{x}_{0}$
	\end{algorithmic} 
\end{algorithm}
\end{figure}

\subsection{High-Fidelity DiffCom: Improving Sampling Efficiency}

In contrast to existing diffusion inverse solvers for vanilla vision applications \cite{chung2022improving, chung2022diffusion, zhu2023denoising}, \emph{DiffCom} confronts the challenge of highly-noisy channel received signal $\bm{y}$ as the measurement and operates the inversion through a complex, highly-nonlinear encoder-function. In that cases, we find these inverse solvers often overfit the forward operator $\mathcal{E}(\cdot)$, i.e., the likelihood score approximation in \eqref{eq:score_likelihood} often leads to a small-enough measurement distance $\mathcal{L}_m$, but by inspecting $\hat{\bm{x}}_{0|t}$, one can easily concludes that it is not aligned with $\bm{y}$. This \emph{ambiguity}, driven by high-power noise and the nonlinear nature of the encoder-function in communication setups, considerably increases the difficulty of achieving stable and efficient diffusion posterior sampling, thereby impacting the quality of the finally decoded images. Addressing this, \emph{DiffCom} necessitates a critical balance: exploring $\bm{x}_{t}$ in proximity to the generative prior, while exploiting the received latent signal to enhance consistency.

\subsubsection{Mitigating Ambiguity with Confirming Constraint}

\begin{figure*}[t]
	\setlength{\abovecaptionskip}{0cm}
	\setlength{\belowcaptionskip}{0cm}
	\centering
	\includegraphics[width=0.85\linewidth]{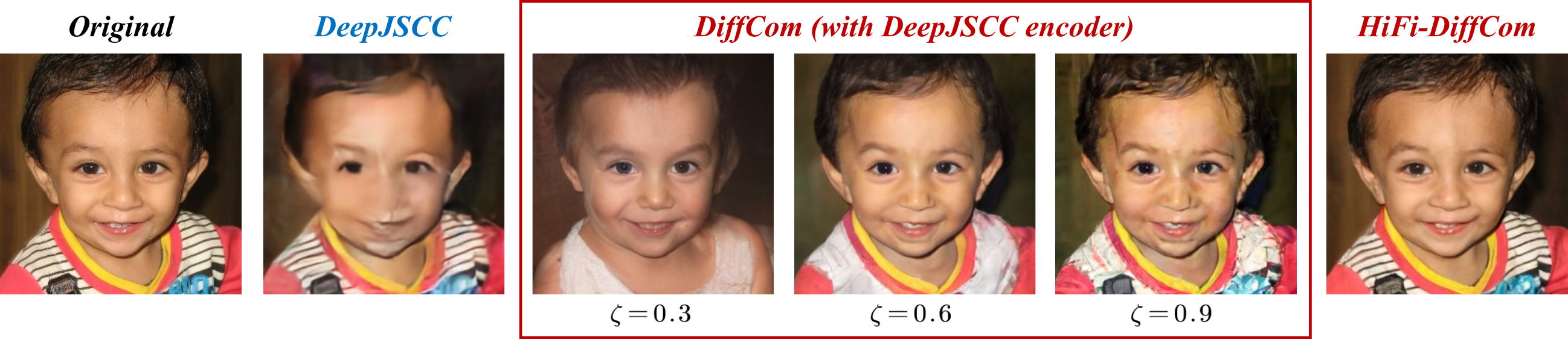}
	\caption{Visual results to demonstrate the effect of hyperparameters $\zeta$, where we evaluate all the schemes under AWGN channel with $\text{CSNR} = 0\text{dB}$.}
	\label{fig:ablation_zeta}
	\vspace{-0em}
\end{figure*}

To address the ambiguity in posterior sampling, we conducted a thorough investigation into the factors contributing to this issue. 
Fig. \ref{fig:ablation_zeta} illustrates the impact of the guidance strength controlling hyperparameter $\zeta$ (we set all $\zeta_t$ to be identical as $\zeta$ in \eqref{eq:score_posterior}) on the visual quality of generated images. 
Our analysis reveals that a lower $\zeta$ value initiates an almost unconditional generation process, producing realistic images that, however, may not faithfully represent the original data. 
Conversely, a higher $\zeta$ value can lead the model to overfit the nonlinear neural encoder-function based on the measurement signal $\bm{y}$. While this may result in sharper images, it often exacerbates noise artifacts, detracting from the overall image fidelity.

Further analysis from Fig. \ref{fig:ablation_zeta} reveals that increasing $\zeta$ initially prioritizes the fitting of high-level content semantics, and consistency in appearance details manifests at higher $\zeta$ values.
However, as $\zeta$ values rise, while consistency in appearance details improves, noise artifacts become prominent in the main body of the content. 
This phenomenon presents a significant challenge in \emph{DiffCom}: \emph{achieving an optimal balance between high-level semantic features and low-level appearance details}, which are manipulated by latent space and source space alignment, respectively.

To address the ambiguity of posterior sampling in standard \emph{DiffCom}, we introduce an enhanced version, ``\emph{HiFi-DiffCom}'', which refines the prior score function with likelihood score functions in both latent space and source space. This modification aims to reduce the overfitting to noisy measurements $\bm{y}$. Specifically, we utilize a paired JSCC decoder $\mathcal{D}$, which is jointly trained with the JSCC encoder $\mathcal{E}$. When provided with $\bm{y}$ as input, the JSCC decoder $\mathcal{D}$ delivers an MSE-optimized reconstruction $\bm{x}_d = \mathcal{D}(\bm{y})$. This output serves as the basis for our newly devised \emph{confirming constraint} term, enhancing the fidelity and robustness of the generated images. 

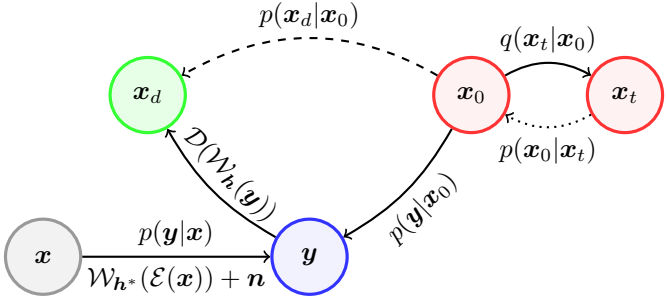
\begin{figure}[t]
	\setlength{\abovecaptionskip}{-0cm}
	\setlength{\belowcaptionskip}{0cm}
	\begin{center}
		\begin{tikzpicture}[
			Nodes/.style={circle, draw=gray!80, fill=gray!10, very thick, minimum size=10mm},
			Nodex/.style={circle, draw=red!80, fill=red!5, very thick, minimum size=10mm},
			Nodey/.style={circle, draw=blue!80, fill=blue!5, very thick, minimum size=10mm},
			Noded/.style={circle, draw=green!80, fill=green!10, very thick, minimum size=10mm}
			]
			\node[Nodey]  (y)  {$\bm y$};
			\node[Nodex]    (x0) [above right= 2of y]   {${\bm x}_0$};
			\node[Nodex]  (xt)   [right= 1of x0] {${\bm x}_t$};
			\node[Noded]    (xd)     [above left = 2of y] {${\bm x}_d$} ;
			\node[Nodes]  (x) [left = 2.5of y]{$\bm x$} ;

			%
			%
			\draw [->, thick] (y) to [out=150,in=-60] node[above,sloped] {$\mathcal{D}(\mathcal{W}_{\bm{h}}(\bm{y}))$} (xd);
			\draw [->, thick] (x) to [out=0,in=180]  node[above] {$p({\bm{y}}|{\bm{x}})$} node[below] {$\mathcal{W}_{\bm{h}^*}(\mathcal{E}(\bm{x})) + \bm{n}$}  (y);
			\draw [->, thick] (x0) to [out=-120,in=30] node[below,sloped] {$p({\bm{y}}|{\bm{x}_0})$}  (y);
			\draw [->, thick] (x0) to [out=30,in=150] node[above] {$q({\bm x}_t|{\bm x}_0)$} (xt);
			\draw [->, thick,dotted] (xt) to [out=210,in=-30] node[below] {$p({\bm x}_0|{\bm x}_t)$} (x0);
			\draw [->, thick,dashed] (x0) to [out=150,in=30] node[above] {$p({\bm x}_d|{\bm x}_0)$} (xd);
			
		\end{tikzpicture}
		\caption{Probabilistic graph of \emph{HiFi-DiffCom}.}
		\label{fig:prob_graph_hifi_diffcom}
	\end{center}
\end{figure}

Under this setting, as shown in Fig. \ref{fig:prob_graph_hifi_diffcom}, the posterior score for diffusion reverse sampling is derived as
\begin{align}
	\nabla_{\bm{x}_t} \log p(\bm{x}_t|\bm{y},\bm{x}_d) & =  \nabla_{\bm{x}_t} \log p(\bm{x}_t) + \nabla_{\bm{x}_t} \log p(\bm{y},\bm{x}_d|\bm{x}_t) \notag \\
	~ &  \approx s_{\bm \theta}(\bm{x}_t, t) + \nabla_{\bm{x}_t} \log p(\bm{y},\bm{x}_d|\bm{x}_t),	\label{eq:score_posterior_hifi}
\end{align}
where the likelihood score for \emph{HiFi-DiffCom} is then formalized as follows by leveraging the proxy $\hat{\bm{x}}_{0|t}$ in \eqref{eq:posterior_mean}:
\begin{align}
	& \nabla_{\bm{x}_t} \log p(\bm{y},\bm{x}_d|\bm{x}_t) = \nabla_{\bm{x}_t} \log p(\bm{y}|\bm{x}_t) \notag \\ 
	& + \nabla_{\bm{x}_t} \log p(\bm{x}_d|\bm{x}_t,\bm{y}) \notag \\ 
	 & \approx  \nabla_{\bm{x}_t} \log p(\bm{y}|\hat{\bm{x}}_{0|t}) + \nabla_{\bm{x}_t} \log p(\bm{x}_d|\hat{\bm{x}}_{0|t}) \notag \\
	& \approx - \zeta_t \nabla_{\bm{x}_t} \underset{\text{for latent space alignment}}{\underbrace{\|\bm{y} - \mathcal{W}_{\bm{h}}(\mathcal{E}(\hat{\bm{x}}_{0|t}))\|_2^2}} +  ~ \notag \\
	& - \gamma_t \nabla_{\bm{x}_t}  \underset{\text{``confirming'' of $\hat{\bm{x}}_{0|t}$ for source space alignment}}{\underbrace{ \Vert \bm{x}_d - 
	\mathcal{D}\left(\mathcal{W}_{\bm{h}}^{-1}\left(\mathcal{W}_{\bm{h}}\left(\mathcal{E}\left(\hat{\bm{x}}_{0|t}\right)\right)\right)\right)\Vert_2^2}}, 
	\label{eq:likelihood_score}
\end{align}
where $\gamma_t$ controls the confirming strength of time step $t$ in the source pixel domain, and the extra confirming term takes into account the effect of nonlinear decoder and wireless channel.

Leveraging JSCC decoder, our extra confirming constraint ensures that the generated sample remains at the manifold of real data, particularly enhancing the reproduction of faithful low-level details consistent with the original. This method substantially surpasses approaches that rely solely on direct pixel domain alignment, such as minimizing $\Vert  \bm{x}_d - \hat{\bm{x}}_{0|t} \Vert_2^2$. Instead, our strategy focuses on aligning the sampled results with the measurements in both the source and latent domains, explicitly considering the nonlinear autoencoder functions and wireless-related operations. A comprehensive numeric comparison is presented in our later detailed ablation study, showcasing the superiority of this method over other variants.

\begin{figure}
	\vspace{-0.5em}
\begin{algorithm}[H]
	\small
	\caption{High-Fidelity Posterior Sampling of DiffCom}\label{alg2}
	\begin{algorithmic}[1] 
		\Require {$\bm{y}$, $\mathcal{E}$, $\mathcal{D}$, $\mathcal{W}_{\bm{h}}$, $\sigma_n^2$, $T$, $\{\zeta_t\}^T_{t=1}$, $\{\gamma_t\}^T_{t=1}$, $\{\sigma_t\}^T_{t=1}$, and $s_{\bm \theta}(\cdot,\cdot)$.}
		\LineComment{Deterministic decoding}
		\State $ \bm{h} = \emph{Channel-Estimation}(\bm{y}_{\text{pilot}})$
		\State $ \bm{x}_d = \mathcal{D}(\mathcal{W}^{-1}_{\bm{h}}(\bm{y}))$
		\LineComment{Timestep initialization as \eqref{eq:time_step_init}}
		\State $T_s = \min \Big(\eta \big(\argmin_{t\in [1,T]} \Vert(1 + \frac{1}{\sigma_n^2})^{{\frac{\rho}{2}} \cdot \tau} - \frac{\overline{\alpha}_t}{(1 - \overline{\alpha}_t)} \Vert_2^2\big), T \Big)$
		\LineComment{Put $\bm{x}_d$ into the matched noise manifold}
		\State $\bm{\epsilon}_{T_s} \sim \mathcal{N} (\bm{0}, \bm{I}_m)$ 
		\State Initialize $\bm{x}_{T_s} = \sqrt{\overline{\alpha}_{T_s}}\bm{x}_d + \sqrt{1 - \overline{\alpha}_{T_s}}\bm{\epsilon}_{T_s}$			
		\For {$t=T_s, \ldots , 1$}
		\State $\hat{\bm{s}} \gets s_{\bm \theta}(\bm{x}_{t},t)$
		\State $\hat{\bm{x}}_{0|t} \gets \frac{1}{\sqrt{\overline{\alpha}}_{t}}\left( \bm{x}_{t} + \left({1-\overline{\alpha}_{t}}\right) \hat{\bm{s}} \right)$
		\State $\bm{\epsilon} \sim \mathcal{N} (\bm{0}, \bm{I}_m)$ \textbf{if} $t > 0$, \textbf{else} $\bm{\epsilon}=\bm{0}$
		\State $\bm{x}_{t-1}^\prime \gets \frac{\sqrt{\alpha_{t}} (1 - \overline{\alpha}_{t-1})}{1 - \overline{\alpha}_{t}}\bm{x}_{t} + \frac{\sqrt{\overline{\alpha}_{t-1}} {\beta}_{t}}{1 - \overline{\alpha}_{t}}\hat{\bm{x}}_{0|t} + \tilde{\sigma}_t \bm{\epsilon}$ 
		\LineComment{Enforcing measurement consistency in latent space}
		\State $\mathcal{L}_m \gets \Vert\bm{y} -\mathcal{W}_{\bm{h}}\left(\mathcal{E}\left(\hat{\bm{x}}_{0|t}\right)\right)\Vert_2^2$
		\LineComment{Confirming low-level details in source space}
		\State $\mathcal{L}_c \gets \Vert\bm{x}_d -\mathcal{D}\left(\mathcal{W}_{\bm{h}}^{-1}\left(\mathcal{W}_{\bm{h}}\left(\mathcal{E}\left(\hat{\bm{x}}_{0|t}\right)\right)\right)\right)\Vert_2^2$
		\LineComment{Finish one step posterior sampling}
		\State $\bm{x}_{t-1} \gets \bm{x}_{t-1}^\prime - \zeta_t \nabla_{\bm{x}_t} \mathcal{L}_m  - \gamma_t \nabla_{\bm{x}_t} \mathcal{L}_c$
		\EndFor
		\State \textbf{return} $\bm{x}_{0}$
	\end{algorithmic} 
\end{algorithm}
\end{figure}

\subsubsection{Accelerate Sampling with Adaptive Initialization}

When the JSCC decoder $\mathcal{D}$ is utilized, it indeed imparts strong prior knowledge, substantially accelerating the diffusion posterior sampling process. Observationally, if the MSE-optimized decoding result $\bm{x}_d$ closely approximates the ground truth, it is advantageous to initiate the reverse sampling from a lightly-noised manifold (corresponding to a smaller timestep) rather than starting from a pure Gaussian distribution $\mathcal{N}(\bm{0}, \bm{I}_m)$ at the maximum timestep $T$. Leveraging this insight, we optimize the use of the channel received signal $\bm{y}$ by starting posterior sampling at an intermediate timestep $T_s$. Specifically, we apply the marginal distribution in \eqref{eq:xt_cond_x0} to appropriately inject noise into $\bm{x}_d$ onto some proper noise manifold:
\begin{equation}
	\bm{x}_{T_s} = \sqrt{\overline{\alpha}_{T_s}}\bm{x}_d + \sqrt{1 - \overline{\alpha}_{T_s}}\bm{\epsilon}_{T_s},~ \bm{\epsilon}_{T_s} \sim \mathcal{N} (\bm{0}, \bm{I}_m).
	\label{eq:init}
\end{equation}

The selection of $T_s$ is crucial and highly dependent on the quality of $\bm{x}_d$. 
Intuitively, $\bm{x}_{T_s}$ should retain more information from $\bm{x}_d$ when its quality is higher. 
Given the pre-defined noise variance schedule $\{\beta_t\}_{t=1}^T$, the SNR of VP diffusion process, referred to as DSNR to differentiate from CSNR, at timestep $t$ is given by $\text{DSNR}(t) = \overline{\alpha}_t / (1 - \overline{\alpha}_t)$. 
To determine an optimal $T_s$, we seek for an alignment between the wireless channel quality and the noisy manifold characteristics of our diffusion model. The end-to-end performance for a test Gaussian source signal transmitted over a wireless channel, characterized by the signal-to-distortion ratio (SDR) at a given CSNR, is evaluated \cite{Saidutta2021joint}. Combining the RD function and the Gaussian channel capacity, the SDR is calculated as $(1 + \frac{1}{\sigma_n^2})^{\frac{\rho}{2}}$ with the CBR $\rho = k/m$ \cite{Saidutta2021joint}. The intermediate timestep $T_s$ is then selected by matching the SDR with the DSNR:
\begin{equation}
	T_s = \min \Bigg(\eta \Big(\argmin_{t\in [1,T]} \Vert(1 + \frac{1}{\sigma_n^2})^{{\frac{\rho}{2}} \cdot \tau} - \frac{\overline{\alpha}_t}{(1 - \overline{\alpha}_t)} \Vert_2^2\Big), T \Bigg),
	\label{eq:time_step_init}
\end{equation}
where $\tau$ is a constant introduced for numerical stability, and $\eta$ is a scaling factor to adjust the total rounds of reverse sampling steps. 
This approach significantly reduces the required reverse sampling steps from $4\times$ to $50\times$, depending on the specific codec $\mathcal{E}$ and $\mathcal{D}$, CBR $\rho$, and CSNR. The whole procedure of the HiFi-DiffCom posterior sampling is given in Algorithm~\ref{alg2}.

\subsection{Blind-DiffCom: Enhancing Sampling Robustness}

In both standard \emph{DiffCom} and \emph{HiFi-DiffCom}, we have demonstrated the utilization of the channel received signal for posterior sampling to establish a high-fidelity generative end-to-end communications framework. 
Typically, the actual channel response $\bm{h}^*$ in wireless operations $\mathcal{W}_{\bm{h}^*}$ is approximated by $\bm{h}$, derived using channel estimation algorithms. 
As an extension, we further propose ``\emph{Blind-DiffCom}'' to address extreme scenarios where channel estimation is unavailable or highly inaccurate, facilitating pilot-free end-to-end communication that enhances system efficiency and robustness.

In the blind scenario, only the prior knowledge about the distribution of $\bm{h}^*$, obtained through channel modeling techniques, is available; the specific channel response remains underestimated or unestimated. 
\emph{Blind-DiffCom} incorporates an additional diffusion model that samples the channel response based on its distribution.
This setup enables the joint estimation of channel parameters and source data in a coarse-to-fine manner by exploiting the interplay between source and channel diffusion models. Similar joint estimation strategies have been successfully employed in blind image restoration techniques, where the forward operation kernel is unknown \cite{chung2023parallel}.

In alignment with the OFDM transmission configurations adopted in \cite{yang2022ofdm}, this paper models a multipath fading channel with $L$ independent paths, each subject to Rayleigh fading, represented as $\bm{h} \sim \mathcal{CN}(\bm{0}, \bm{\sigma}_h^2 \bm{I}_L)$.
The power for each path, $\sigma_{h,l}^2$ for $l=0,1, \cdots, L-1$, is characterized by an exponential decay profile: $\sigma_{h,l}^2 = c_l e^{-\frac{l}{r}}$, where $c_l$ is a normalization coefficient ensuring $\sum_{l=0}^{l-1} \sigma_{h,l}^2 = 1$, and $r$ is the decay rate or time delay constant \cite{yang2022ofdm}. In the context of \emph{Blind-DiffCom}, our optimization objective is redefined to jointly estimate the source data and the channel vector as follows:
\begin{equation}
	\argmin_{\bm{x}_0, \bm{h}_0}{\mathcal{P}_{\bm{x}}(\bm{x}_0) + \mathcal{P}_{\bm{h}}(\bm{h}_0) + \zeta \|\bm{y} - \mathcal{W}_{\bm{h}_0}(\mathcal{E}(\bm{x}_0))\|_2^2},
	\label{eq:blind_hqs}
\end{equation}
where $\bm{h}_0$ denotes the estimated channel impulse response in time domain, and $\mathcal{P}_{\bm{x}}$, $\mathcal{P}_{\bm{h}}$ are regularization functions for the source image and the channel vector, respectively.

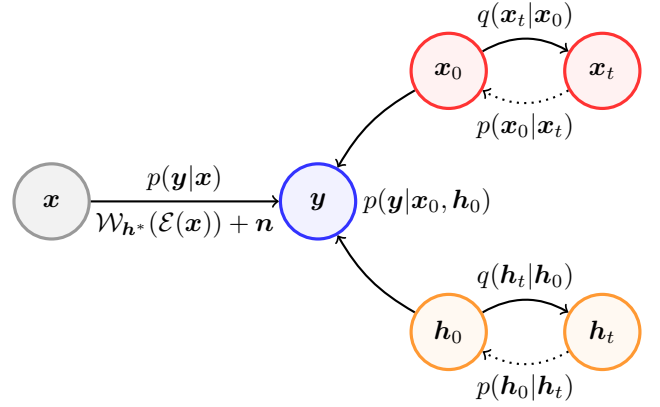
\begin{figure}[t]
	\setlength{\abovecaptionskip}{-0cm}
	\setlength{\belowcaptionskip}{0cm}
	\begin{center}
		\begin{tikzpicture}[
			Nodes/.style={circle, draw=gray!80, fill=gray!10, very thick, minimum size=10mm},
			Nodex/.style={circle, draw=red!80, fill=red!5, very thick, minimum size=10mm},
			Nodey/.style={circle, draw=blue!80, fill=blue!5, very thick, minimum size=10mm},
			Nodeh/.style={circle, draw=orange!80, fill=orange!5, very thick, minimum size=10mm}
			]
			\node[Nodes]  (x) {$\bm x$}  ;
			\node[Nodey]  (y)  [right=2.5of x]  {$\bm y$}  ;
			\node[Nodex]    (x0)     [above right=of y] {${\bm x}_0$} ;
			\node[Nodex]  (xt)   [right=1of x0] {${\bm x}_t$};
			\node[Nodeh]    (h0)     [below right=of y] {${\bm h}_0$};
			\node[Nodeh]    (ht)     [right=1of h0] {${\bm h}_t$};
			\filldraw[black] (5,0) circle (0pt) node {$p({\bm y}|{\bm x}_0,{\bm h}_0)$};
			%
			\draw [->, thick] (x0) to [out=-150,in=60]   (y);
			\draw [->, thick] (x) to [out=0,in=180] node[above] {$p({\bm{y}}|{\bm{x}})$} node[below] {$\mathcal{W}_{\bm{h}^*}(\mathcal{E}(\bm{x})) + \bm{n}$}  (y);
			\draw [->, thick] (x0) to [out=30,in=150] node[above] {$q({\bm x}_t|{\bm x}_0)$} (xt);
			\draw [->, thick,dotted] (xt) to [out=210,in=-30] node[below,sloped] {$p({\bm x}_0|{\bm x}_t)$} (x0);
			\draw [->, thick] (h0) to [out=150,in=-60]  (y);
			\draw [->, thick] (h0) to [out=30,in=150] node[above] {$q({\bm h}_t|{\bm h}_0)$} (ht);
			\draw [->, thick,dotted] (ht) to [out=210,in=-30] node[below,sloped] {$p({\bm h}_0|{\bm h}_t)$} (h0);
		\end{tikzpicture}
		\caption{Probabilistic graph of \emph{Blind-DiffCom}.}
		\label{fig:prob_graph_blind_diffcom}
	\end{center}
\end{figure}

\begin{figure}[t]
	\setlength{\abovecaptionskip}{-0cm}
	\setlength{\belowcaptionskip}{0cm}
	\centering
	\includegraphics[width=0.98\linewidth]{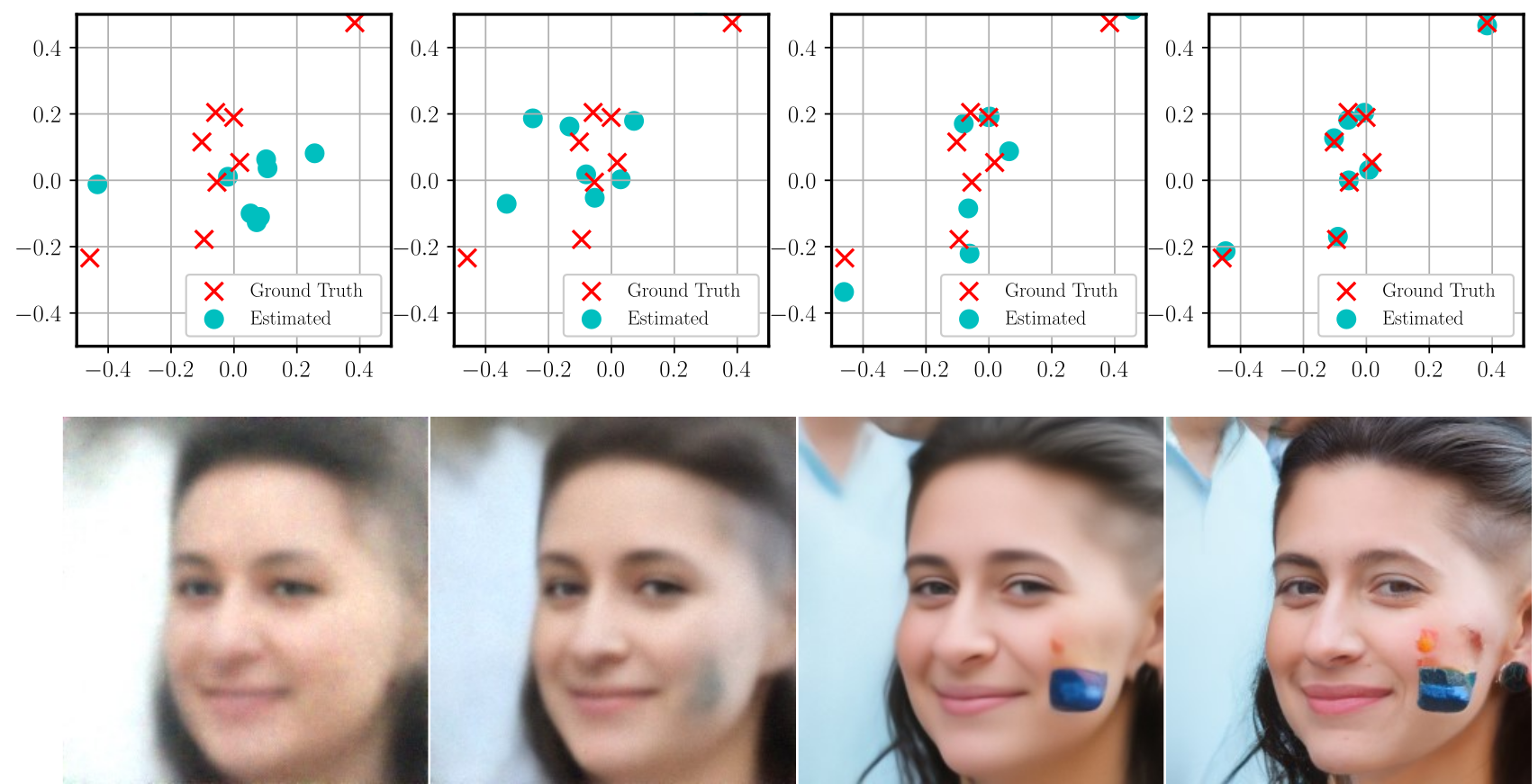}
	\caption{Progress of channel estimation-free transmission over a multipath fading channel with $L=8$. Top row: trajectories of estimated $\bm{h}_{t|0}$ at $t=1000$, $750$, $500$, and $0$, respectively. Bottom row: evolution of $\bm{x}_{t|0}$ over time.
	}
	\label{fig:blind_diffcom}
\end{figure}

We solve \eqref{eq:blind_hqs} iteratively in a coarse-to-fine manner through posterior sampling.
As demonstrated in Fig. \ref{fig:prob_graph_blind_diffcom}, since $\bm{x}_0$ and $\bm{h}_0$ are typically independent, the posterior distribution can be factorized as
\begin{equation}
	p(\bm{x}_0, \bm{h}_0 | \bm{y}) \propto p(\bm{y} | \bm{x}_0, \bm{h}_0)p(\bm{x}_0)p(\bm{h}_0),
	\label{eq:bayes_joint}
\end{equation}
where the likelihood term $p(\bm{y} | \bm{x}_0, \bm{h}_0)$ is represented as
\begin{equation}
	p(\bm{y} | \bm{x}_0, \bm{h}_0) = \mathcal{N}(\bm{y}; \mathcal{W}_{\bm{h}_0}(\mathcal{E}(\bm{x}_0)), \sigma_n^2 \bm{I}_k).
	\label{eq:blind_measurement_model}
\end{equation}
As a natural extension of \emph{DiffCom}, \emph{Blind-DiffCom} constructs the posterior score for sampling source as follows: 
\begin{align}
	\nabla_{\bm{x}_t} \log p(\bm{x}_t, \bm{h}_t|\bm{y}) &\approx s_{\bm \theta}(\bm{x}_t, t) - ~\notag \\
	& {\zeta_t} \nabla_{\bm{x}_t} {\|\bm{y} - \mathcal{W}_{\hat{\bm{h}}_{0|t}}(\mathcal{E}(\hat{\bm{x}}_{0|t}))\|_2^2}.
	\label{eq:posterior_score_x}
\end{align}
Consistent with the formulation in \eqref{eq:score_likelihood}, the likelihood score is tied to the source posterior mean $\hat{\bm{x}}_{0|t}$, but involves a wireless-related operator parameterized by the estimated posterior mean $\mathcal{W}_{\hat{\bm{h}}_{0|t}}$, where the channel posterior is computed as \eqref{eq:posterior_mean}. For channel variable $\bm{h}_t$, since we know the analytical expression for its distribution, $p(\bm{h}_t) = \mathcal{CN}(\bm{h}_t; \bm{0}, \bm{\sigma}_h^2 \bm{I}_L)$, its score function can be analytically derived as $\nabla_{\bm{h}_t} \log p(\bm{h}_t) = -{\bm{h}_{t}} / {\bm{\sigma}_h^2}$ for any timestep $t$. The resulting expression for the channel posterior score is
\begin{align}
	\nabla_{\bm{h}_t} \log p(\bm{x}_t, \bm{h}_t|\bm{y}) = -\frac{\bm{h}_{t}}{\bm{\sigma}_h^2} - {\zeta_t} \nabla_{\bm{h}_t} {\|\bm{y} - \mathcal{W}_{\hat{\bm{h}}_{0|t}}(\mathcal{E}(\hat{\bm{x}}_{0|t}))\|_2^2}.
	\label{eq:posterior_score_h}
\end{align}
This configuration allows for the parallel updating of $\bm{h}_{t}$ and $\bm{x}_{t}$. 
The steps of the \emph{Blind-DiffCom} posterior sampling process are outlined in Algorithm \ref{alg3}. 
We illustrate the joint estimation process visually in Fig. \ref{fig:blind_diffcom}, conducted under a multipath fading channel with CSNR = 10dB and $L=8$. 
The results clearly show that both the source image and the channel response begin with coarse estimations and progressively refine toward their ground truths over successive timesteps.
Given the analytical form of the channel diffusion model, it is noteworthy that \emph{Blind-DiffCom} incurs only minimal additional computational costs. This efficiency significantly enhances its applicability and generalization capabilities for in-the-wild scenarios.

\begin{figure}
	\begin{algorithm}[H]
		\small
		\caption{Blind Posterior Sampling of DiffCom}\label{alg3}
		\begin{algorithmic}[1] 
			\Require {$\bm{y}$, $\mathcal{E}$, $\mathcal{W}_{\bm{h}}$, $T$, $\{\zeta_t\}^T_{t=1}$, $\{\sigma_t\}^T_{t=1}$, and $s_{\bm \theta}(\cdot,\cdot)$.}
			\State Initialize $\bm{x}_T \sim \mathcal{N} (\bm{0}, \bm{I}_m)$, $\bm{h}_T \sim \mathcal{CN} (\bm{0}, \bm{I}_L)$
			\For {$t=T, \ldots , 1$}
			\State $\hat{\bm{s}} \gets s_{\bm \theta}(\bm{x}_{t},t)$
			\State $\hat{\bm{x}}_{0|t} \gets \frac{1}{\sqrt{\overline{\alpha}_{t}}}\left( \bm{x}_{t} + \left({1-\overline{\alpha}_{t}}\right) \hat{\bm{s}} \right)$
			\State $\hat{\bm{h}}_{0|t} \gets \frac{1}{\sqrt{\overline{\alpha}_{t}}}\left( \bm{h}_{t} + \left({1-\overline{\alpha}_{t}}\right) \frac{\bm{h}_{t}}{\bm{\sigma}_h^2} \right)$
			\State $\bm{\epsilon}_x \sim \mathcal{N} (\bm{0}, \bm{I}_m)$ \textbf{if} $t > 1$, \textbf{else} $\bm{\epsilon}_x=\bm{0}$
			\State $\bm{\epsilon}_h \sim \mathcal{N} (\bm{0}, \bm{I}_L)$ \textbf{if} $t > 1$, \textbf{else} $\bm{\epsilon}_h=\bm{0}$
			\LineComment{Source and channel diffusion ancestral sampling}
			\State $\bm{x}_{t-1}^\prime \gets \frac{\sqrt{\alpha_{t}} (1 - \overline{\alpha}_{t-1})}{1 - \overline{\alpha}_{t}}\bm{x}_{t} + \frac{\sqrt{\overline{\alpha}_{t-1}} {\beta}_{t}}{1 - \overline{\alpha}_{t}}\hat{\bm{x}}_{0|t} + \tilde{\sigma}_t \bm{\epsilon}_x$ 
			\State $\bm{h}_{t-1}^\prime \gets \frac{\sqrt{\alpha_{t}} (1 - \overline{\alpha}_{t-1})}{1 - \overline{\alpha}_{t}}\bm{h}_{t} + \frac{\sqrt{\overline{\alpha}_{t-1}} {\beta}_{t}}{1 - \overline{\alpha}_{t}}\hat{\bm{h}}_{0|t} + \tilde{\sigma}_t \bm{\epsilon}_h$ 
			\LineComment{Enforcing data consistency through estimated $\hat{\bm{h}}_{0|t}$}
			\State $\mathcal{L} \gets \Vert\bm{y} -\mathcal{W}_{\hat{\bm{h}}_{0|t}}\left(\mathcal{E}\left(\hat{\bm{x}}_{0|t}\right)\right)\Vert_2^2$
			\LineComment{Finish one step guided posterior sampling}
			\State $\bm{x}_{t-1} \gets \bm{x}_{t-1}^\prime - \zeta_t \nabla_{\bm{x}_t} \mathcal{L}$
			\State $\bm{h}_{t-1} \gets \bm{h}_{t-1}^\prime - \zeta_t \nabla_{\bm{h}_t} \mathcal{L}$
			\EndFor
			\State \textbf{return} $\bm{x}_{0}$, $\bm{h}_{0}$
		\end{algorithmic} 
	\end{algorithm}
\end{figure}

\section{Experimental Results}

\subsection{Implementation Details}

\subsubsection{Dataset and Metrics}

We conduct experiments on FFHQ $256 \times 256$ \cite{karras2019style} and ImageNet $256 \times 256$ \cite{deng2009imagenet} datasets. 
For each dataset, we evaluate $100$ hold-out validation images.
We use three types of image quality assessment (IQA) methods to provide a thorough performance evaluation:
(a) pixel-level distortion metrics, such as PSNR, which focus on the quantitative accuracy of the pixel values; 
(b) low-level consistency metrics, such as LPIPS \cite{lpips} and DISTS \cite{ding2020iqa}, which are deep feature-based distances developed to model the Human Visual System (HVS);
(c) high-level realism metrics, such as Fréchet Inception Distance (FID) \cite{heusel2017gans}, which are no-reference metrics focusing on the distributional alignment, widely used to access the quality of generated~images. We use ``$\downarrow$'' and ``$\uparrow$'' to mark that lower or higher metrics represent better quality.

\subsubsection{Diffusion Models and End-to-end Transmission Models}

Our methodology is adaptable to a wide range of diffusion and differentiable JSCC models. For all experiments, we utilize pre-trained DDPM models from \cite{nichol2021improved} and \cite{choi2021ilvr} on the ImageNet and FFHQ datasets, respectively. We retain their predefined noise variance schedules, where $\beta$ decays linearly from an initial value of $0.02$ to $0.0001$ over 1,000 discretized timesteps.

{For JSCC models, we integrate three representative RD-optimized image JSCC methods: DeepJSCC \cite{djscc}, SwinJSCC \cite{yang2024swinjscc}, and the improved NTSCC \cite{wang2023improved}. Specifically, we implement the compatible NTSCC+ model \cite{wang2023improved}, which supports SNR-adaptive and fine-grained rate adjustments within a single model. }
We use parentheses to indicate the specific JSCC encoder paired with our \emph{DiffCom}, such as ``\emph{DiffCom} (NTSCC)'', ``\emph{DiffCom} (SwinJSCC)'' and ``\emph{DiffCom} (DeepJSCC)''.
By default, for \emph{DiffCom}, we use a constant guidance strength $\zeta = 0.6$. 
For \emph{HiFi-DiffCom}, we set $\zeta=0.3$ and $\gamma=0.3$, with the starting timestep $T_s$ determined using $\tau=20$ and $\eta=1$ as \eqref{eq:time_step_init}.
While further performance improvements could be achieved by finely tuning these hyperparameters, we primarily focus on demonstrating the efficacy of proposed methods.

\subsubsection{Configurations for OFDM Transmission} 

In our evaluation of end-to-end communication under multipath fading conditions, we implement an OFDM transmission configured with $L_{\text{ifft}}=256$ sub-carriers and add $L_{\text{cp}}=10$ CP symbols to each frame. 
Each OFDM sub-carrier carries $N_s= \lceil k / L_{\text{fft}} \rceil$ data symbols and $N_p=1$ pilot symbol along the time axis. 
The multipath channel model consists of $L=8$ independent paths with a constant time decay parameter of $r=4$. 
For channel estimation and equalization, we employ the linear minimum mean square error (LMMSE) technique. 
To enhance robustness against unexpected channel fading, we also incorporate an interleaver to shuffle the JSCC coded symbols $\bm{z}$ prior to transmission through the OFDM system.

\subsubsection{Comparison Schemes}

To evaluate the proposed method, we conducted comparative analyses with established transmission techniques across several categories. 
For separation-based image transmission schemes, our benchmarks include mainstream image codecs, both handcrafted and learned, paired with 5G LDPC channel coding (code length 4096) and digital modulation in line with the 3GPP TS 38.214 standard. 
Handcrafted codecs assessed include VTM (intra coding for VVC, the state-of-the-art codec) and BPG (compliant with HEVC intra coding). 
MSE-optimized Neural Image Codecs (NICs) Hyperprior \cite{balle2018variational} and GMM-Attn \cite{cheng2020image} are also considered for comparison.
{ Moreover, we further compare our approach against previous state-of-the-art perceptually-oriented codecs: including HiFiC \cite{mentzer2020high}, CDC \cite{yang2024lossy}, and MS-ILLM\cite{muckley2023improving}.
These codecs incorporate MSE, LPIPS, and adversarial loss into their optimization criteria.
Specifically, CDC \cite{yang2024lossy} is a recent perceptual codec that employs conditional diffusion model, while MS-ILLM \cite{muckley2023improving} represents the previous state-of-the-art in perceptual image compression. It is important to acknowledge that our comparisons are conducted under low CBR regions and various SNRs. In some instances, the performance for some competitive schemes are missing because they do not provide pretrained models that support these configurations.}

In the category of end-to-end optimized image transmission methods, we compare against DeepJSCC \cite{djscc, xu2021wireless}, SwinJSCC \cite{yang2024swinjscc}, NTSCC \cite{wang2023improved}, and perceptually optimized DeepJSCC (PDJSCC) \cite{wang2022perceptual}. 
All neural codecs were tested using their respective open-source implementations with pretrained weights to maintain consistency with published results.

\subsection{Rate-Distortion-Perception Performance Comparison}

\subsubsection{Quantitative Results}

\begin{figure*}[t]
	\setlength{\abovecaptionskip}{-0.cm}
	\setlength{\belowcaptionskip}{-0.cm}
	\centering
	\includegraphics[width=\linewidth]{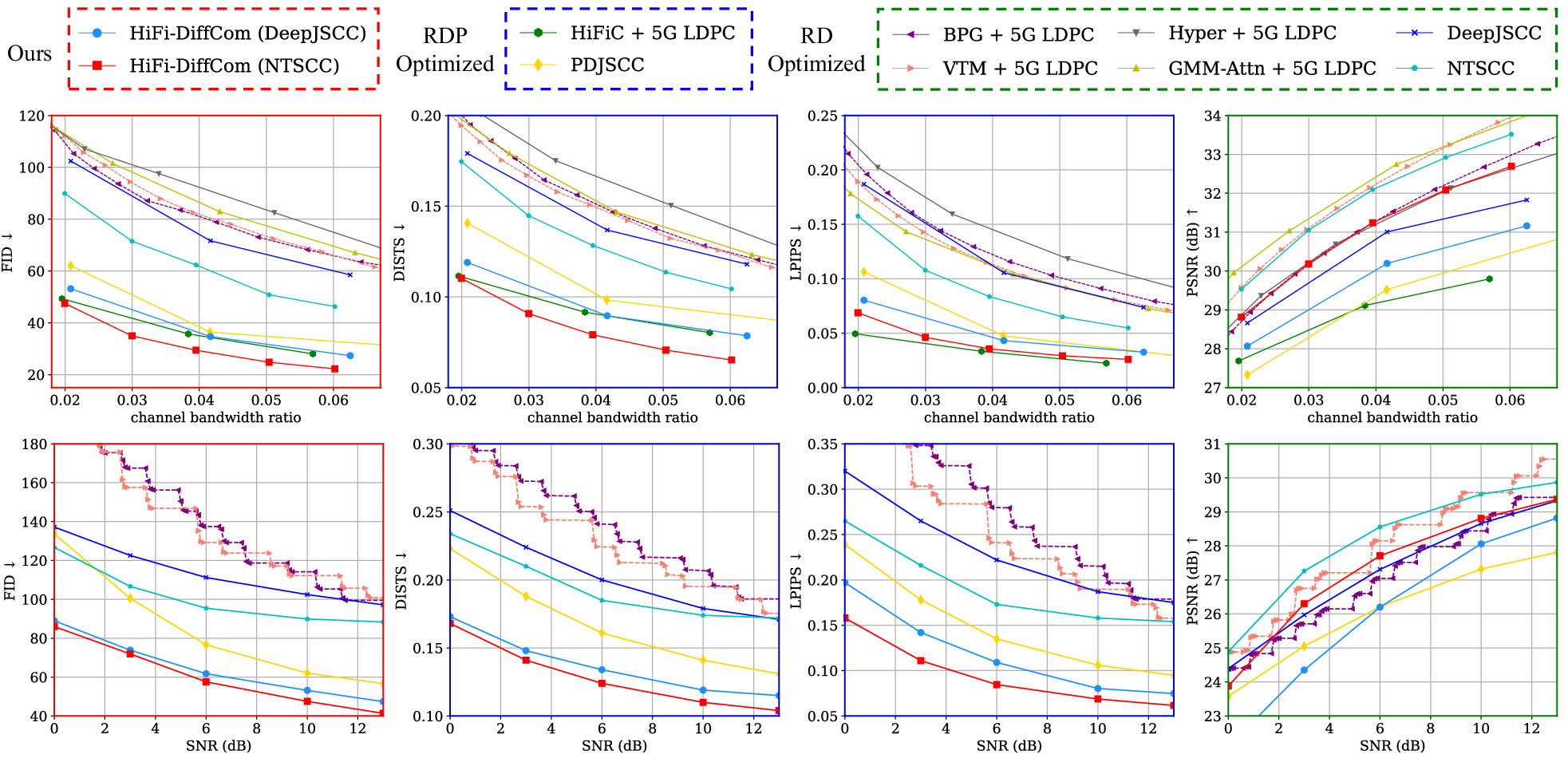}
	\caption{Comparisons of wireless image transmission methods across various distortion and fidelity metrics, tested on FFHQ testset under AWGN channel. Metrics are categorized into realism (red frames), consistency (blue frames), and distortion (green frames). First row: reconstruction quality vs. channel bandwidth ratio (CBR) tested under CSNR = 10dB; Second row: reconstruction quality vs. channel SNR with a very low CBR of approximately $1/48$.}
	\label{fig:RDP_curve}
\end{figure*}

\begin{table}[t]
	\scriptsize
	\centering
	\caption{
		RDP results on ImageNet under AWGN channel CSNR = 10dB.
	}
	\tabcolsep=0.08cm
	\resizebox{\linewidth}{!}{
		\begin{tabular}{m{2.7cm}<{\centering}m{0.8cm}<{\centering}m{1.2cm}<{\centering}m{0.8cm}<{\centering}m{0.8cm}<{\centering}m{0.8cm}<{\centering}m{0.8cm}<{\centering}m{0.8cm}}
			\toprule
			{{ImageNet}, CSNR=10dB} & \multicolumn{2}{c}{Distortion} & \multicolumn{2}{c}{Consistency} & \multicolumn{3}{c}{Realism} \\
			\cmidrule(lr){1-1}
			\cmidrule(lr){2-3}
			\cmidrule(lr){4-5}
			\cmidrule(lr){6-8}
			{Method} & {PSNR$\uparrow$} & {MS-SSIM$\uparrow$} & {LPIPS$\downarrow$} & {DISTS$\downarrow$} & {NIQE$\downarrow$} & {HIQA$\uparrow$} & {FID$\downarrow$} \\
			\midrule
			
			VTM + 5G LDPC~ &  \underline{30.05} & 0.96 & 0.202 & 0.206 & 10.58 & 0.52 & 123.7\\
			NTC \cite{cheng2020image} + 5G LDPC & \textbf{30.33} & \underline{0.97} & 0.188 & 0.215 & 9.94 & 0.63 & 135.9\\
			HiFiC \cite{mentzer2020high} + 5G LDPC & 26.39 & 0.96 & {0.065} & {0.132} & \underline{6.15} & 0.55 & {53.9}\\
			DeepJSCC \cite{xu2021wireless} & 28.55 & 0.97 &  0.211 & 0.200 & 6.85 & \underline{0.66} & 164.9 \\
			SwinJSCC \cite{yang2024swinjscc} & 29.90 & 0.97 & 0.171 & 0.182 & 8.64 & 0.53 & 88.3\\
			NTSCC \cite{wang2023improved}~ & 29.88 & \textbf{0.97} & {0.166} & {0.188} & {6.85}  & \textbf{0.66} & {89.4}\\
			PDJSCC \cite{wang2022perceptual}~ & 26.00 & 0.96 & 0.123 & 0.161 & 6.73 & 0.49 & 69.4 \\
			MS-ILLM \cite{muckley2023improving}~ & 28.28 & 0.96 & \textbf{0.051} & \textbf{0.114} & \textbf{5.97} & 0.61 & \underline{46.4} \\
			\midrule
			\emph{HiFi-DiffCom} (DeepJSCC) & 27.96 &  0.95 &  0.142 & 0.143 &  6.22 & 0.60 & {52.6}\\
			\emph{HiFi-DiffCom} (NTSCC) & 29.02 & 0.96 & \underline{0.088} & \underline{0.117} & \underline{6.05} & {0.64} & \textbf{43.2} \\
			\bottomrule
		\end{tabular}
	}
	\label{tab:imagenet_10dB}
\end{table}

\begin{table}[t]
	\centering
	\caption{
		RDP results on ImageNet under AWGN channel CSNR = 0dB.
	}
	\tabcolsep=0.08cm
	\resizebox{\linewidth}{!}{
		\begin{tabular}{m{2.7cm}<{\centering}m{0.8cm}<{\centering}m{1.4cm}<{\centering}m{0.8cm}<{\centering}m{0.8cm}<{\centering}m{0.8cm}<{\centering}m{0.8cm}<{\centering}m{0.8cm}}
			\toprule
			{{ImageNet}, CSNR=0dB} & \multicolumn{2}{c}{Distortion} & \multicolumn{2}{c}{Consistency} & \multicolumn{3}{c}{Realism} \\
			\cmidrule(lr){1-1}
			\cmidrule(lr){2-3}
			\cmidrule(lr){4-5}
			\cmidrule(lr){6-8}
			{Method} & {PSNR$\uparrow$} & {MS-SSIM$\uparrow$} & {LPIPS$\downarrow$} & {DISTS$\downarrow$} & {NIQE$\downarrow$} & {HIQA$\uparrow$} & {FID$\downarrow$} \\
			\midrule
			VTM + 5G LDPC & \underline{25.12} & 0.88 & 0.421 & 0.312 & 12.21 & 0.30 & 242.6 \\
			DeepJSCC \cite{xu2021wireless} & 24.62 & 0.89 & 0.358 & 0.277 & 6.85 & \underline{0.66} & 213.9\\
			SwinJSCC \cite{yang2024swinjscc} & \textbf{25.53} & \underline{0.90} &  0.293 & 0.251 & 9.16 &  0.42 & 152.1\\
			NTSCC \cite{wang2023improved}~ & 24.89 & \textbf{0.90} & 0.284 & 0.249 & 6.85 & \textbf{0.66} & 166.7 \\
			PDJSCC \cite{wang2022perceptual}~ & 22.89 & 0.88 & \underline{0.266} & {0.245} & 6.58 & 0.35 & 183.5 \\
			\midrule
			\scriptsize \emph{HiFi-DiffCom} (DeepJSCC) & 22.07 & 0.83 &  0.283 & \underline{0.203} & \underline{5.54} & 0.60 & \underline{113.8}\\
			\scriptsize \emph{HiFi-DiffCom} (NTSCC) & 23.83 & 0.88 & \textbf{0.211} & \textbf{0.191} & \textbf{5.50} & 0.60 & \textbf{106.5}\\
			\bottomrule
		\end{tabular}
	}
	\label{tab:imagenet_0dB}
\end{table}

Fig. \ref{fig:RDP_curve} presents a benchmark of advanced wireless image transmission methods on the FFHQ dataset under an AWGN channel. The first row evaluates rate-fidelity curves at a fixed $10$dB CSNR, while the second row examines reconstruction quality across various CSNRs with a CBR of $1/48$.
Moreover, in Tables \ref{tab:imagenet_10dB} and \ref{tab:imagenet_0dB}, we expand our comparative analysis using a larger-scale ImageNet database, assessing image quality under two CSNRs of 10dB and 0dB, respectively, with a very low CBR of approximately $1/48$. We report both full-reference distortion metrics, such as MS-SSIM, and no-reference realism metrics, including NIQE \cite{mittal2012making} and HIQA \cite{su2020blindly}, to provide a comprehensive assessment encompassing both classical and advanced modern quality indicators.

{ The results reveal that while some existing methods excel in specific categories such as distortion, consistency, or realism, they often underperform in others. In contrast, our \emph{HiFi-DiffCom} consistently dominates in terms of the realism metric FID while exhibiting competitive performance in consistency and signal distortion metrics. Compared with CDC + 5G LDPC and MS-ILLM + 5G LDPC, our approach achieves better FID scores and significantly higher PSNR values.
While many schemes outperform ours in consistency metrics, these results are expected since HiFiC, CDC, and MS-ILLM utilize LPIPS as a loss function during training. 
Our approach is optimized for distribution-preserving (i.e., realism) reconstructions, which do not specifically aim to achieve state-of-the-art instance-level consistency metrics like LPIPS.
	
Furthermore, when evaluating across variable CSNRs, as shown in the second row of Fig. \ref{fig:RDP_curve}, separation-based codecs (e.g., NIC + LDPC) suffer from noticeable cliff or leveling effects when channel conditions fall below or exceed the levels anticipated by the channel code. In contrast, our method and JSCC codecs provide smooth performance transitions.
Moreover, because our approach shares the same channel-received signal with RD-optimized JSCC codecs, we are capable of decoding both optimal perceptual quality and optimal MSE simultaneously.
}

\makeatletter
\renewcommand{\@thesubfigure}{\hskip\subfiglabelskip}
\makeatother

\begin{figure*}[t]
	\setlength{\abovecaptionskip}{-0.cm}
	\setlength{\belowcaptionskip}{-0.cm}
	\begin{subtable}
		\centering
		\small
		\begin{tabular}{m{0.2\textwidth}<{\centering}m{0.15\textwidth}<{\centering}m{0.18\textwidth}<{\centering}m{0.17\textwidth}<{\centering}m{0.17\textwidth}<{\centering}}
			Original image & PDJSCC & HiFiC + 5G LDPC & MS-ILLM + 5G LDPC & HiFi-DiffCom
		\end{tabular}
	\end{subtable}
	\vspace{-0.5em}
	\begin{center}
		\subfigure[CBR $\rho$] {\includegraphics[width=0.19\textwidth]{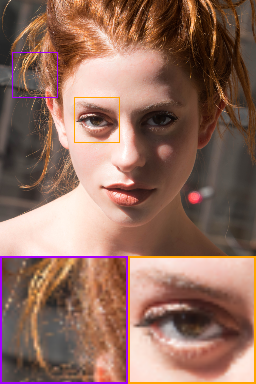}}
		\subfigure[$0.0208$ {(\color{red}$+23\%$})] {\includegraphics[width=0.19\textwidth]{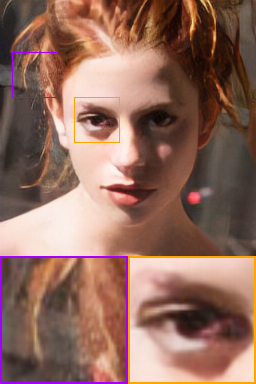}}
		\subfigure[$0.0203$ ({\color{red}$+21\%$})] {\includegraphics[width=0.19\textwidth]{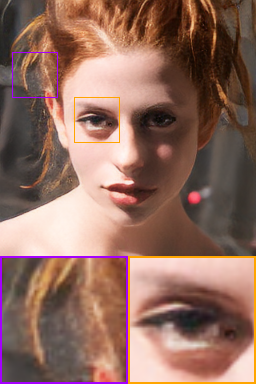}}
		\subfigure[$0.0206$ {(\color{red}$+23\%$})] {\includegraphics[width=0.19\textwidth]{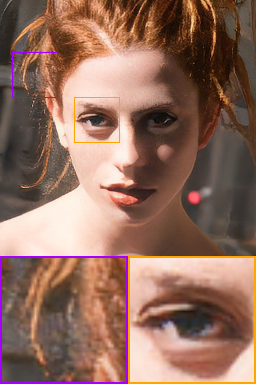}}
		\subfigure[$0.0168$ ({\color{blue}$0\%$})] {\includegraphics[width=0.19\textwidth]{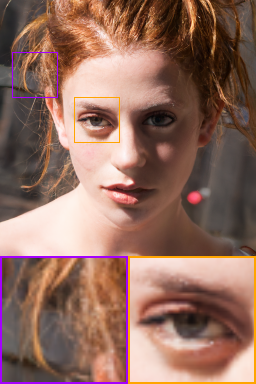}}
		\caption{{Qualitative results under AWGN channel with CSNR = 10dB. $\rho$ denotes the channel bandwidth ratio. The red numbers indicate the extra percentage of bandwidth cost compared to the our \emph{HiFi-DiffCom}. Please zoom in for a better view.}}
		\label{fig:visual}
	\end{center}
	\vspace{0em}
\end{figure*}

\subsubsection{Qualitative Results}

\begin{figure*}[t]
	\setlength{\abovecaptionskip}{-0.cm}
	\setlength{\belowcaptionskip}{-0.cm}
	\centering
	\includegraphics[width=0.98\linewidth]{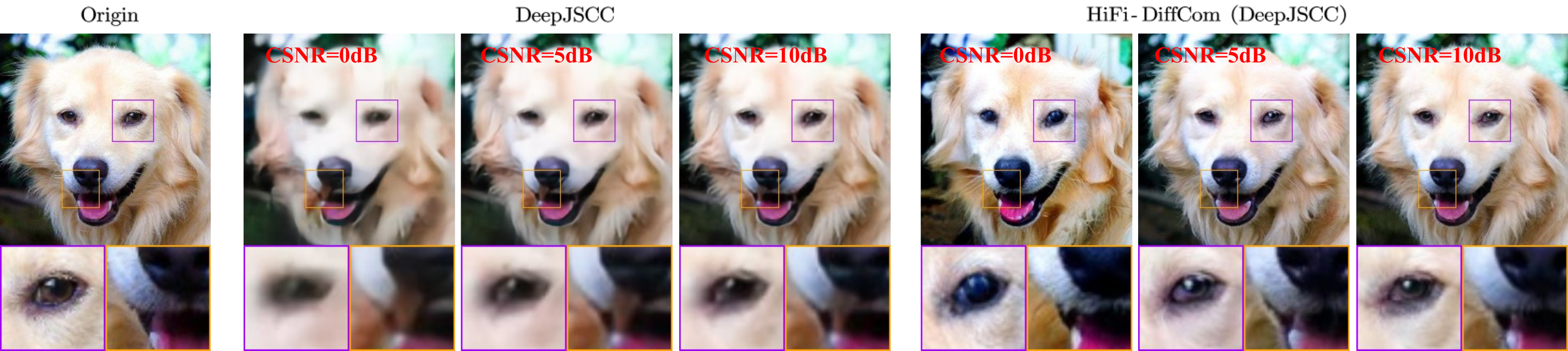}
	\caption{A visual example illustrating how generative transmission methods preserve perfect realism, while exhibiting graceful consistency degradation in details when CSNRs decrease. }
	\label{fig:consistency_degradation}
	\vspace{-0em}
\end{figure*}

In Fig. \ref{fig:visual}, we present qualitative visual comparisons of our approach with advanced perceptual-oriented image transmission schemes. 
Apparently, our \emph{HiFi-DiffCom} performs visually more pleasing with lower channel bandwidth cost. 
Fig. \ref{fig:consistency_degradation} illustrates the reconstructions by ``DeepJSCC'' and our ``\emph{HiFi-DiffCom} (DeepJSCC)'' across various CSNRs from 10dB to 0dB at a very low CBR of $\rho=1/48$. 
Under low CSNR conditions, traditional DeepJSCC reconstructions exhibit decreased realism, characterized by blurring and inconsistent details due to artifact introduction. 
In contrast, our \emph{HiFi-DiffCom} system always produces visually appealing results with faithful details, where the consistency of these details decreases gracefully as CSNR decreases.

\subsection{Generalization and Robustness Comparison}

\begin{figure}[t]
	\setlength{\abovecaptionskip}{-0.cm}
	\setlength{\belowcaptionskip}{-0.cm}
	\centering
	{\includegraphics[width=0.98\linewidth]{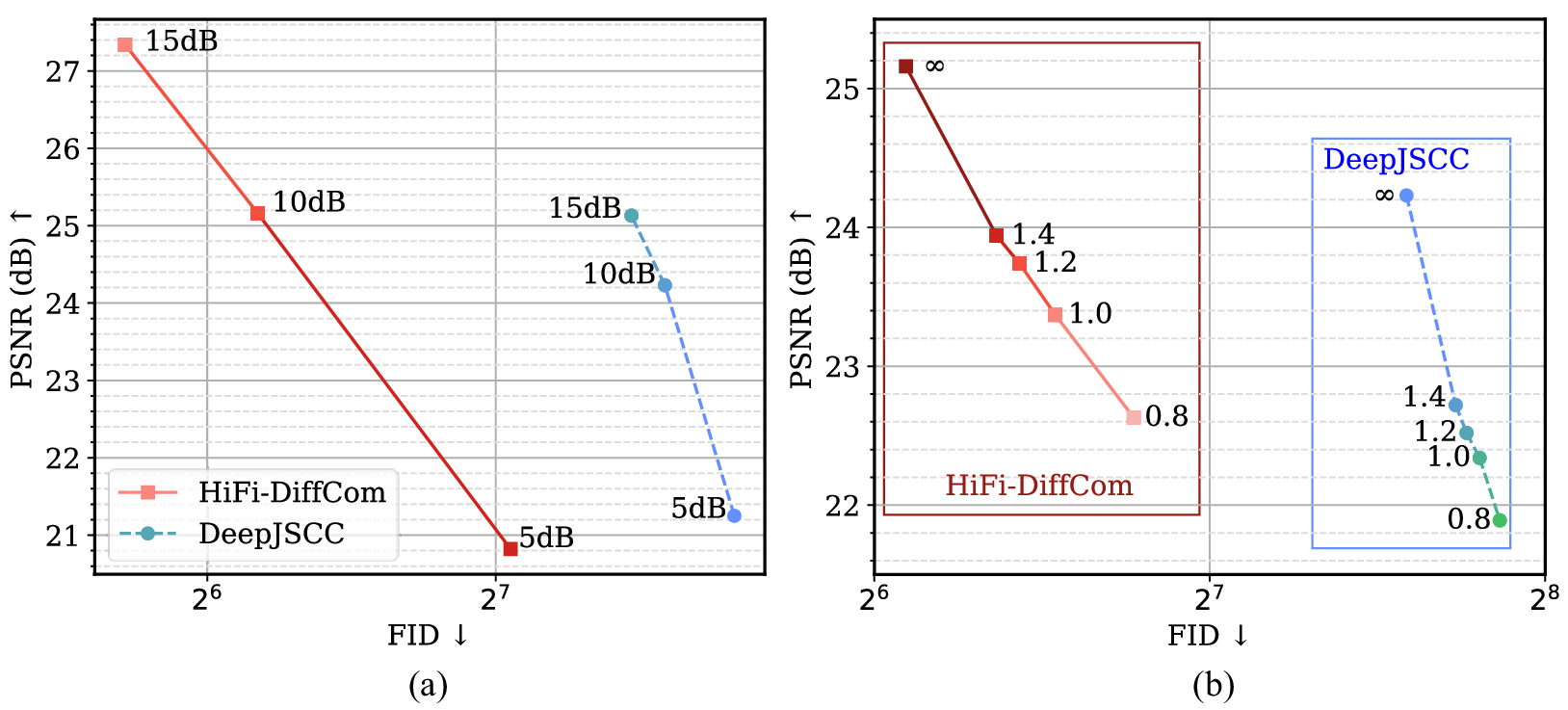}}
	\caption{Distortion (PSNR) vs. realism (FID) under multipath fading channel, evaluated (a) across multiple CSNRs, and (b) at different  clipping ratios with a fixed 10dB CSNR ($\infty$ denotes no clipping). Each marker includes the corresponding CSNR or clipping ratio for clarity.}	\label{fig:fading_and_clipping}
\end{figure}

\begin{figure*}[t]
	\setlength{\abovecaptionskip}{-0.cm}
	\setlength{\belowcaptionskip}{-0.cm}
	\centering
	{\includegraphics[width=0.8\linewidth]{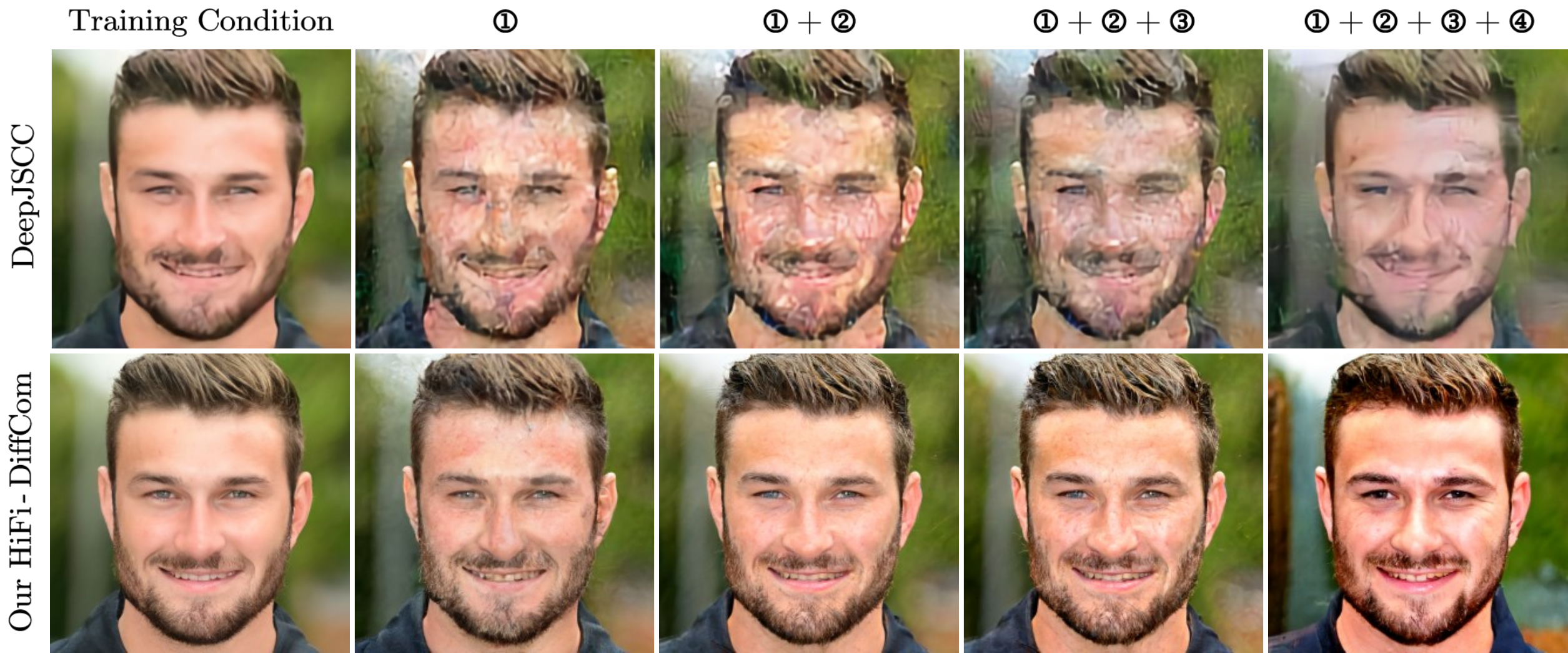}}
	\caption{A visual comparison illustrating the impact of several unexpected transmission degradations: \ding{172} unseen channel fading, \ding{173} PAPR reduction, \ding{174} with ISI (removed CP symbols), and \ding{175} very low CSNR (0dB).}
	\label{fig:generalization}
\end{figure*}

In this subsection, we evaluate the generalization capability and robustness of the proposed diffusion posterior sampling algorithms. The baseline model adopted is a DeepJSCC codec, pre-trained under an AWGN channel and unexposed to fading, clipping, or interference during training. 
We introduce various unexpected wireless transmission-related degradations to its encoded JSCC codewords to assess and compare the reconstruction quality between our \emph{DiffCom} and the traditional DeepJSCC deterministic decoder.

Results under multipath fading channel with MMSE channel estimation across multiple CSNRs is presented in Fig. \ref{fig:fading_and_clipping}(a). 
We report distortion vs. realism for a fixed CBR $\rho=1/48$, measured by PSNR and FID, where the optimal point is at the top left.
It is clear that \emph{HiFi-DiffCom} outperforms DeepJSCC by a large margin in terms of FID metric, and also exhibits PSNR gain for high CSNR points.

Our methods support flexible formulations of the forward operator $\mathcal{W}_{\bm{h}^*}$. 
As an example, we integrate a PAPR reduction module after OFDM modulation to clip OFDM signals. 
This module serves as a peak clipping function, limiting the amplitude of each complex-valued transmission symbol to no more than $c\sqrt{P_s}$, where $c$ is the clipping ratio and $P_s$ is the average transmission signal power \cite{yang2022ofdm}. 
Results in Fig. \ref{fig:fading_and_clipping}(b) confirm that, decoding from the same received signal, \emph{HiFi-DiffCom} maintains superior FID and PSNR versus DeepJSCC.

To verify robustness, Fig. \ref{fig:generalization} depicts the exposure of DeepJSCC codewords to a series of unexpected degradations, decoding the received signals using both our proposed stochastic posterior sampling paradigm and the deterministic DeepJSCC decoder. 
The results demonstrate that the RD-optimized deterministic decoder produces blurry reconstructions even under trained conditions and is particularly vulnerable to transmission perturbations, highlighting its inferior fidelity and inferior robustness. 
Conversely, \emph{HiFi-DiffCom} maintains high-quality image details, effectively counteracting the effects of in-the-wild transmission degradations.

\begin{figure}[t]
	\setlength{\abovecaptionskip}{-0.cm}
	\setlength{\belowcaptionskip}{-0.cm}
	\centering
	\includegraphics[width=0.98\linewidth]{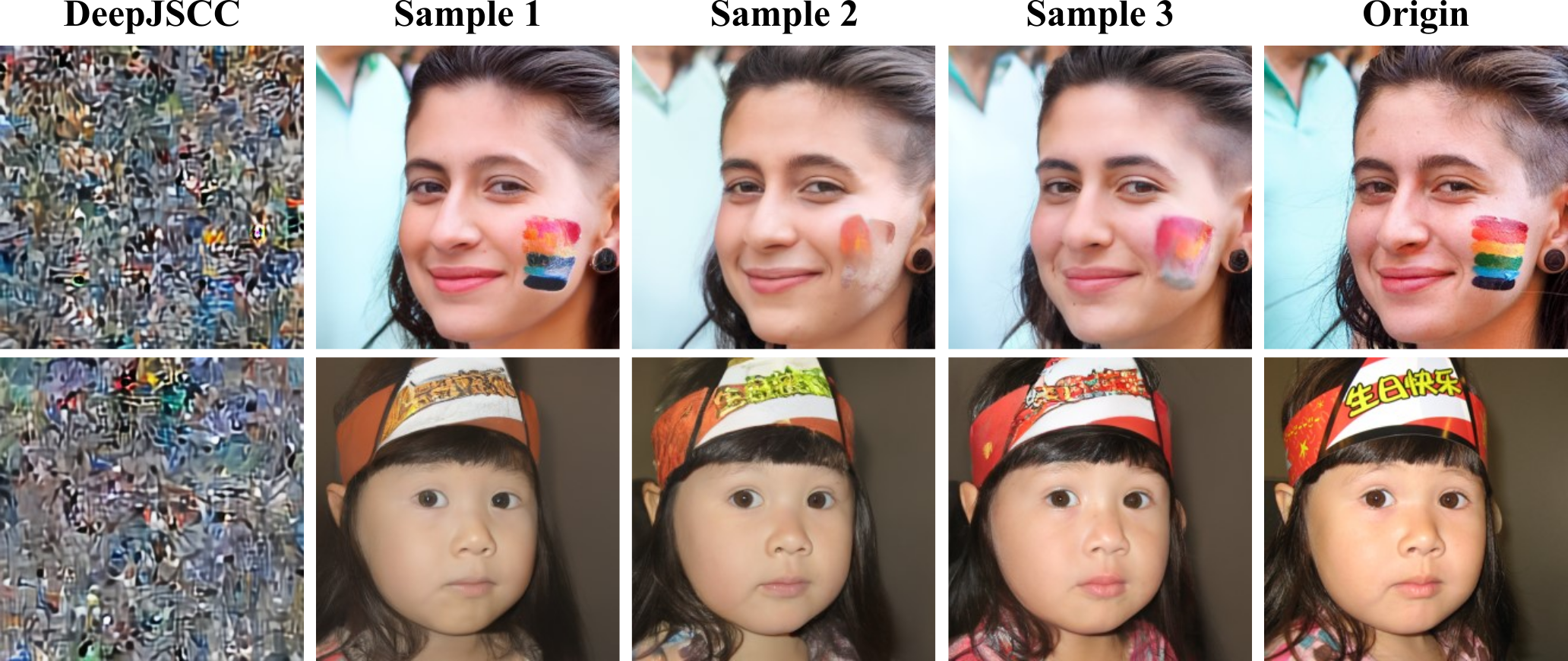}
	\caption{Reconstructions under multipath fading channel without channel estimation ($L=8$, CSNR = 10dB). }
	\label{fig:ce_free}
\end{figure}

\begin{table}[t] 
	\caption{Performance of Blind-DiffCom for pilot-free transmission.}  
	\tabcolsep=0.15cm
	\centering
	\resizebox{\linewidth}{!}{%
		\begin{threeparttable}[b]
			\begin{tabular}{lcccccccc}
				\toprule    
				{Method} & {$\zeta_{\bm{x}}$} & {$\zeta_{\bm{h}}$}  & {Success ratio} & {$d_{\bm{h}}$} & {PSNR$\uparrow$} & {LPIPS$\downarrow$} & {DISTS$\downarrow$} &{FID$\downarrow$} \\
				\midrule
				{DeepJSCC} & / & / & {0\%} & / & 8.23 & 0.828 & 0.473 & 376.3\\
				\cmidrule(lr){1-9}
				\multirow{5}{*}{\shortstack{Blind-DiffCom \\ without CE}} & 0.3 & 0.1 & \textbf{85\%} & 0.265 & 18.19 & 0.297 & 0.216 & 105.5 \\
				& 0.3 & 0.2 & {73\%} & 0.204 & 17.42 & 0.323 & 0.230 & 115.3 \\
				& 0.3 & 0.3 & {58\%} & 0.193 & 17.35 & 0.320 & 0.228 & 118.3 \\
				\cmidrule(lr){2-9}
				& 0.6 & 0.1 & {35\%} & 0.089 & \textbf{22.66} & 0.234 & \textbf{0.188} & 106.6 \\
				& 0.9 & 0.1 & {43\%} & \textbf{0.041} & 20.18 & \textbf{0.228} & 0.190 & \textbf{98.5} \\
				\bottomrule
			\end{tabular}
		\end{threeparttable}
	}
	\label{tab:blind_case}
	\vspace{-1em}
\end{table}

In a challenging scenario without channel estimation (CE), most existing end-to-end transmission methods underperform or fail due to their heavy reliance on accurate CE. 
In Table \ref{tab:blind_case} and Fig. \ref{fig:ce_free}, we evaluate the performance on the FFHQ dataset under a multipath channel with $10$dB CSNR. 
The visual results in Fig. \ref{fig:ce_free} demonstrate that DeepJSCC fails in the absence of CE, yielding garbled images. 
In contrast, our \emph{Blind-DiffCom} robustly produces diverse and accurate reconstructions. 
We also observe that the joint estimation process of the source image and channel response does not always succeed.
In Table \ref{tab:blind_case}, we report the success ratio of \emph{Blind-DiffCom} across 100 different test images, taking into account the guidance strengths $\zeta_{\bm{x}}$ and $\zeta_{\bm{h}}$. 
We calculate the averaged quality metrics and channel estimation precision in terms of $d_{\bm{h}} = \|\bm{h} - \bm{h}^*\|_2^2$ only for meaningful reconstructions, identified by PSNR $\geq$ 15dB. 
Empirical observations suggest that while higher guidance strengths may improve reconstruction quality, they also reduce the likelihood of success. 
Lower guidance strengths, conversely, yield meaningful reconstructions more frequently but with some slightly reduced quality. One effective strategy is to iteratively perform joint posterior sampling on the same received signal to enhance the chances of obtaining valuable reconstructions. 
The potential of \emph{Blind-DiffCom} for pilot-free communication scenarios highlights an important avenue for future research.

\subsection{Ablation Study}

In this subsection, we conduct several ablation studies to validate our key design choices. 
By default, our proposed \emph{DiffCom} methods are evaluated paired with DeepJSCC encoder, which is trained under AWGN channel, with CBR $\rho=1/48$. 

\begin{table}[t]
	\caption{Ablation studies on the design components of HiFi-DiffCom.}  
	\setlength{\tabcolsep}{2pt}
	\centering
	\resizebox{\linewidth}{!}{%
		\begin{tabular}{cccccccccc}
			\toprule        
			\multicolumn{2}{c}{Components} & \multicolumn{4}{c}{CSNR = 10dB} & \multicolumn{4}{c}{CSNR = 0dB}   \\
			\cmidrule(lr){1-2}
			\cmidrule(lr){3-6}
			\cmidrule(lr){7-10}
			{Adapt. init.}  & {Conf. term} & {PSNR$\uparrow$} & {LPIPS$\downarrow$}& {DISTS$\downarrow$} &{FID$\downarrow$}  &{PSNR$\uparrow$}& {LPIPS$\downarrow$} & {DISTS$\downarrow$}& {FID$\downarrow$}  \\
			\midrule
			\xmark &  \xmark & 19.98 &  0.161 & 0.160 & 71.1 & 18.1 & 0.239 & 0.202 & 97.2  \\  
			\cmark & \xmark & 27.77 & 0.091 & 0.121 & \underline{53.9} & 20.42 & 0.262 & 0.211  & 104.2\\        
			\xmark &  \cmark & 27.63 & \underline{0.088} & \underline{0.120} & 54.2 & \underline{23.34} & \textbf{0.181} & \textbf{0.164} & \textbf{81.2}  \\
			\rowcolor{lightgray}
			\cmark & \cmark  & \underline{28.06} & \textbf{0.080} & \textbf{0.119} & \textbf{53.2} & 22.1 & \underline{0.197} & \underline{0.173} & \underline{89.0} \\
			\midrule
			\multicolumn{2}{c}{DeepJSCC} & \textbf{28.66}  & 0.187 & 0.179 & 102.4 & \textbf{24.39} & {0.320} & 0.251 & 137.2 \\
			\bottomrule
		\end{tabular}
	} 
	\label{tab:ablation_design_component}
\end{table}

\begin{figure*}[t]
	\setlength{\abovecaptionskip}{-0.cm}
	\setlength{\belowcaptionskip}{-0.cm}
	\centering
	\includegraphics[width=0.9 \linewidth]{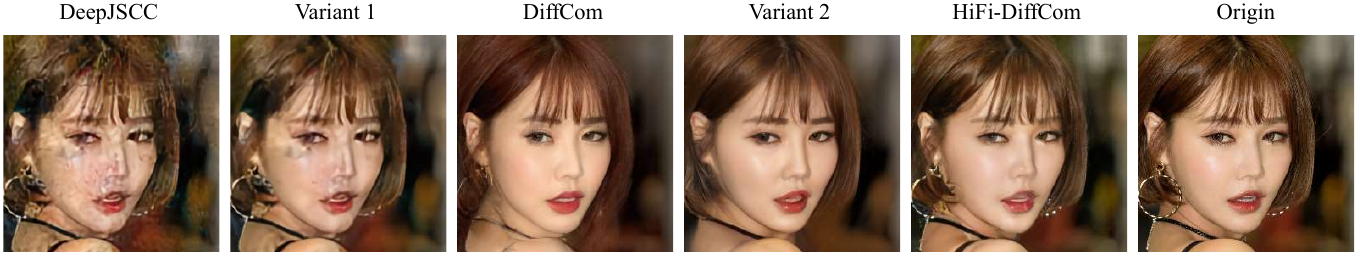}
	\caption{Reconstructions under a multipath fading channel. DeepJSCC, which was trained under an AWGN channel, displays severe artifacts. In contrast, decoding from the same channel received signal, our proposed \emph{DiffCom} and \emph{HiFi-DiffCom} maintain realistic reconstructions.}
	\label{fig:ablation_variants}
\end{figure*}

\subsubsection{Contributions of the Components of HiFi-DiffCom}

In Table \ref{tab:ablation_design_component} and Fig. \ref{fig:ablation_variants}, we dissect the key design components of \emph{HiFi-DiffCom} to assess the impact of the proposed confirming constraint and accelerating sampling strategy. 
The confirming loss significantly improves sample quality across all evaluated metrics. 
Additionally, the adaptive initialization strategy not only accelerates the generation process by reducing the number of required sampling steps, but also influences the reconstruction performance based on the quality of $\bm{x}_d$ used for initialization. 
When combined, these components enable \emph{HiFi-DiffCom} (highlighted in Table \ref{tab:ablation_design_component}) to produce high-quality reconstructions in substantially fewer timesteps.

\begin{table}[t]
	\caption{Ablation studies on the latent guidance strength $\zeta$.}  
	\centering
	\resizebox{\linewidth}{!}{%
		\begin{threeparttable}[b]
			\begin{tabular}{lccccccc}
				\toprule   
				{Method}  & {$\zeta$} & {$\mathcal{L}_m$} & {PSNR$\uparrow$} & {LPIPS$\downarrow$} & {DISTS$\downarrow$} &{FID$\downarrow$} \\
				\midrule
				{DeepJSCC}  & {/} & {/} & \textbf{28.66} & {0.187} & {0.179} &{102.4} \\
				\multirow{4}{*}{\emph{DiffCom}}  & 0.3 & {38.09}& {17.48} & 0.266 & 0.214 & 91.3 \\
				& 0.6 & 44.37 & {18.81} & {0.203} & {0.180} & {78.4} \\   
				& 0.9 & 48.10 & {19.98} & \underline{0.161}  & \underline{0.160} & \underline{71.0} \\             
				& 1.2 & \underline{27.66} & {16.29}  & 0.263 & 0.249 & 122.3 \\
				\midrule
				\rowcolor{lightgray}
				\emph{HiFi-DiffCom} ($\gamma$=0.3) & 0.3  & \textbf{4.60} & \underline{28.06}  & \textbf{0.080} & \textbf{0.119} & \textbf{53.2} \\
				\bottomrule
			\end{tabular}
		\end{threeparttable}
	}
	\label{tab:ablation_zeta}
\end{table}

{
\subsubsection{Impact of Latent Guidance Strength}

As intuitively illustrated in Fig. \ref{fig:ablation_zeta}, we investigate the effect of varying the latent guidance strength $\zeta$. To quantitatively assess this impact, we present ablation studies in Table \ref{tab:ablation_zeta}. The results indicate that for \emph{DiffCom}, the measurement distance $\mathcal{L}_m$ in the latent space does not accurately reflect the fidelity of the reconstructions. In fact, when the operator comprises a deep neural network, closer alignment with the measurement may paradoxically degrade sample quality. This behavior substantially differs from that observed in existing inverse solvers, where closer alignment typically leads to better fidelity, as their operators are usually shallow linear or nonlinear transforms \cite{choi2021ilvr, chung2022diffusion, zhu2023denoising}. Results show that \emph{HiFi-DiffCom} effectively mitigates overfitting to the neural network operator, as evidenced by a substantially lower $\mathcal{L}_m$ and improved FID score.
}

\begin{table*}[t]
	\caption{Ablation studies on different formulations of likelihood score.}  
	\setlength{\tabcolsep}{3pt}
	\centering
		\begin{tabular}{ccccccccccccc}
			\toprule   
			
			\multirow{2}{*}{Method} & \multirow{2}{*}{Initialization}  & \multirow{2}{*}{Constraint}
			& \multirow{2}{*}{\shortstack{\\Likelihood \\ approxiamation}}   
			& \multicolumn{4}{c}{AWGN channel} & \multicolumn{4}{c}{Multipath Fading channel} \\
			\cmidrule(lr){5-8}
			\cmidrule(lr){9-12}  
			{} & {} & {} & & {$T_s$ $\downarrow$} & {PSNR$\uparrow$} & {LPIPS$\downarrow$} & {FID$\downarrow$} & {$T_s$ $\downarrow$} & {PSNR$\uparrow$} & {LPIPS$\downarrow$} &  {FID$\downarrow$} \\
			\midrule
			DeepJSCC & / & / & / & / & \underline{28.66} & 0.187  & 102.4 & / & 24.23 & 0.311  & 192.2\\
			Variant 1 & $\mathcal{N} (\bm{0}, \bm{I}_m)$ & $\bm{x}_d$ & $\Vert\bm{x}_d -\mathcal{D}(\mathcal{W}_{\bm{h}}^{-1}(\mathcal{W}_{\bm{h}}(\mathcal{E}(\hat{\bm{x}}_{0|t}))))\Vert_2^2$ & 1000 & \textbf{28.69} & 0.194  & 100.4 &  1000 &\underline{24.79} & 0.232 & 126.6  \\
			\emph{DiffCom} & $\mathcal{N} (\bm{0}, \bm{I}_m)$ &  $\bm{y}$ & $\|\bm{y} - \mathcal{W}_{\bm{h}}(\mathcal{E}(\hat{\bm{x}}_{0|t}))\|_2^2$ &  1000 &19.98 & \underline{0.161} & \underline{71.0} &  1000 &18.18 & 0.227  & 105.5\\
			Variant 2 & $\mathcal{N} (\bm{0}, \bm{I}_m)$ &  $\bm{x}_d$, $\bm{y}$ & $\|\bm{y} - \mathcal{W}_{\bm{h}}(\mathcal{E}(\hat{\bm{x}}_{0|t}))\|_2^2 + \Vert \bm{x}_{d} - \hat{\bm{x}}_{0|t} \Vert_2^2$ & 1000 &26.88 & 0.161  & 76.3 &  1000 &23.20 & \underline{0.200}  & \underline{82.8}\\            
			\rowcolor{lightgray}
			\emph{HiFi-DiffCom} & Eq. \eqref{eq:init} &  $\bm{x}_d$, $\bm{y}$ &  Eq. \eqref{eq:likelihood_score} & \textbf{171} & {28.06} & \textbf{0.080}  & \textbf{53.2} & \textbf{330} & \textbf{25.16}  & \textbf{0.143} & \textbf{68.3} \\
			\bottomrule
		\end{tabular}
	\label{tab:ablation_likelihood}
\end{table*}

{
\subsubsection{Formulations of Likelihood Score Function}

How to effectively guide the posterior sampling matters in \emph{DiffCom} framework.
Given that the prior term comes from pre-trained unconditional diffusion model, we examine the impact of different formulations on the likelihood score, as detailed in Table \ref{tab:ablation_likelihood} and intuitively visualized in Fig. \ref{fig:ablation_variants}.
Specifically, we investigate two variant methods, referred to as Variant 1 and Variant 2, by evaluating their performance under both AWGN channel conditions (aligned with the base codec training) and multipath fading channel conditions (representing unexpected degradations for the base codec).
For Variant 1, we use the reconstructed image $\bm{x}_{d}$ as condition, and implement the gradient step in Algorithm \ref{alg1} as $\nabla_{\bm{x}_t} \Vert\bm{x}_d -\mathcal{D}(\mathcal{W}_{\bm{h}}^{-1}(\mathcal{W}_{\bm{h}}(\mathcal{E}(\hat{\bm{x}}_{0|t}))))\Vert_2^2$. 
For Variant 2, we aim to check the case where the confirming loss term in Algorithm \ref{alg2} is computed as raw pixel-domain distortion $\Vert \bm{x}_{d} - \hat{\bm{x}}_{0|t} \Vert_2^2$.

Compared to the basic codec DeepJSCC, the results show that Variant 1 does not significantly improve realism under the AWGN channel (FID 100.4 vs. 102.4) and only moderately reduces artifacts under the fading channel (see Fig. \ref{fig:ablation_variants}). 
In contrast, \emph{DiffCom} produces more realistic images in both channel conditions, as evidenced by much lower FID scores than Variant 1 (71.0 vs. 100.4 for the AWGN case and 105.5 vs. 126.6 for the fading case). 
This demonstrates the efficacy of our approach, which leverages the raw channel-received signal for guidance rather than relying on post-processing already blurred or heavily corrupted images, particularly in fading scenarios.
Moreover, our \emph{HiFi-DiffCom} outperforms Variant 2 while using fewer timesteps, demonstrating its greater effectiveness by explicitly accounting for the nonlinear autoencoder and wireless channel operations, rather than relying solely on pixel domain alignment.
}

\begin{table}[t]
	\caption{Ablation studies on decoding steps.}  
	\centering
	\resizebox{\linewidth}{!}{%
		\begin{threeparttable}[b]
			\begin{tabular}{lcccccccc}
				\toprule   
				{Method} & {$\eta$} & {Steps} & {Dec. time} & {PSNR$\uparrow$} & {LPIPS$\downarrow$} & {DISTS$\downarrow$} &{FID$\downarrow$} \\
				\midrule
				{DeepJSCC} & {/} & {/} & {$0.01$s} & \textbf{28.66} & {0.187} & {0.179} &{102.4} \\
				\shortstack{\emph{DiffCom}} & {/} & 1000 & $\sim 50$s & {17.48} & 0.266 & 0.214 & 91.3 \\
				\midrule
				\multirow{6}{*}{\shortstack{\emph{HiFi-DiffCom}}} & 0.05 & 9 & 0.67s & \underline{28.59} & 0.154 & 0.160 & {83.7} \\   
				& 0.1 & 17 & 1.3s & 28.55 & {0.141}  & {0.153} & {75.8} \\             
				& 0.2 & 34 & 2.0s  & 28.39 & 0.136  & 0.146 & 71.1 \\
				& 0.5 & {86} & 5.2s & {28.09}  & {0.111} & {0.130} & {59.7} \\
				& 1.0 & {171} & 10.8s & {28.06}  & \textbf{0.080} & \textbf{0.119} & \textbf{53.2} \\
				& 2.0 & {342} & 22.3s & {27.78}  & \underline{0.096} & \underline{0.122} & \underline{53.9} \\
				\bottomrule
			\end{tabular}
		\end{threeparttable}
	}
	\label{tab:ablation_steps}
\end{table}

{
	\subsubsection{The effect of sampling steps} 
	In Table \ref{tab:ablation_steps}, we examine how reconstruction quality varies with different numbers of decoding steps. By adjusting $\eta$ in Eq. \eqref{eq:time_step_init}, \emph{HiFi-DiffCom} can initiate the reverse diffusion process at an intermediate timestep, thereby reducing the total number of decoding steps required. 
	Our findings indicate that incorporating more decoding steps is crucial for achieving better reconstruction quality in terms of both consistency and realism metrics, and the best FID score is achieved for $\eta=1.0$. Additionally, our method achieves multiple distortion-realism trade-offs using the same channel-received signal. The receiver can decide how much detail to generate, balancing between a reconstruction close to the input (higher PSNR) and one that looks more realistic.
	
	Furthermore, in Table \ref{tab:ablation_steps}, we report the runtime measured on an RTX 4090 GPU, averaged over 100 images from the FFHQ dataset. Our methods have the same encoding time ($\sim$0.01s) as the basic codec DeepJSCC but increase the decoding time from $\sim$0.01s to $\sim$10.8s due to the requirement of gradient ascent during testing. 
	Compared with \emph{DiffCom}, \emph{HiFi-DiffCom} reduces the number of decoding steps by $\sim$ 100x (from 1,000 to 9 steps), while achieving a better FID (83.7 vs. 91.3). This results in a significant speedup, decreasing the decoding time from $\sim$ 50s to 0.67s.
	However, when compared with conditional generative codecs (e.g., HiFiC and MS-ILLM paired with 5G LDPC), our approach run much slower.
	The decoding process of HiFiC and MS-ILLM + 5G LDPC only takes $\sim$0.1s, and the decoding time for the CDC + 5G LDPC method (implemented with x-parameterization models) is 1.12s on our platform. 
		
	As a novel decoding paradigm for end-to-end communication systems, we emphasize that, \emph{DiffCom} series is the preferred method, especially when the basic JSCC codec presents unconvincing reconstructions or breaks down due to challenging communication environments.
	Although the current decoding speed hinders real-time applications, our method will benefit from future researches into speeding up diffusion models, possible directions including fast SDE solvers \cite{lu2022dpm}, progressive distillation \cite{salimansprogressive}, and cache reuse mechanism \cite{ma2024deepcache}, which will be a focus of our future work.
}

\section{Conclusion}

This paper has introduced \emph{DiffCom}, a pioneering framework designed to catalyze the paradigm shift to generative communications. 
In this new paradigm, the channel received signal is utilized not merely as input to a deterministic decoder but as a powerful guidance for generative posterior sampling. 
We have demonstrated that a posterior sampler, by leveraging the received signal as a fine-grained condition for diffusion posterior sampling, can achieve perfect perceptual quality while retaining faithful details consistent with the ground truth at the same time, thus overcoming the limitations inherent in traditional deterministic decoders. 
This approach not only enhances the reliability of generative communication systems but also significantly improves their robustness and generalization capabilities.
Furthermore, we have developed two variants of \emph{DiffCom}: \emph{HiFi-DiffCom}, which enhances the efficiency of posterior sampling, and \emph{Blind-DiffCom}, which caters to scenarios lacking precise or even without channel estimations.
These advancements verifies that robust, stochastic posterior sampling algorithms can provide superior data recovery under extreme wireless transmission conditions.
This work takes an important step towards fostering generative perceptual communication and lays foundational insights for future developments in semantic communication design.

\ifCLASSOPTIONcaptionsoff
\newpage
\fi

\bibliographystyle{IEEEtran}
\bibliography{Ref}
\end{document}